\documentclass[aps,prd,notitlepage,nofootinbib,showpacs,superscriptaddress,groupedaddress]{revtex4-1}  
\usepackage{dcolumn}
\usepackage{epsfig}
\usepackage{graphicx,rotating,booktabs}
\usepackage{amsfonts}
\usepackage{amssymb}
\usepackage{mathtools}
\usepackage{slashed}
\usepackage{bm}
\usepackage{latexsym}
\usepackage{amsmath}
\usepackage{hyperref}
\usepackage{color}
\usepackage{enumitem}
\usepackage[margin=2.5cm]{geometry}
\usepackage[verbose]{placeins}

\usepackage{blindtext}

\newcommand{\beq}{\begin{equation}}
\newcommand{\eeq}{\end{equation}}
\newcommand{\bea}{\begin{eqnarray}}
\newcommand{\eea}{\end{eqnarray}}

\begin{document}
\setlength{\tabcolsep}{6pt}
\renewcommand{\arraystretch}{1.2}


\title{The Spectrum of Darkonium in the Sun}

\author{Chris Kouvaris}
\email{kouvaris@cp3.sdu.dk}
\author{Kasper Lang\ae ble}
\email{langaeble@cp3.sdu.dk}
\author{Niklas Gr\o nlund Nielsen}
\email{ngnielsen@cp3.sdu.dk}
\affiliation{CP$^3$-Origins, Centre for Cosmology and Particle Physics Phenomenology University of Southern Denmark, Campusvej 55, 5230 Odense M, Denmark}

\begin{abstract}
Dark matter that gets captured in the Sun may form positronium-like bound states if it self-interacts via light dark photons. 
In this case, dark matter  can either annihilate to dark photons or recombine in bound states which subsequently also decay to dark photons. The fraction of the dark photons that leave the Sun without decaying to Standard Model particles have a characteristic energy spectrum which is a mixture of the direct annihilation process, the decays of ortho- and para- bound states and the recombination process. The ultimate decay of these dark photons to positron-electron pairs (via kinetic mixing) outside  the Sun creates a distinct signal that can either identify or set strict constraints on dark photon models.
\\[.1cm]
{\footnotesize  \it Preprint: CP3-Origins-2016-029 DNRF90}
\end{abstract}


\maketitle

\section{Introduction}
Dark matter (DM) is approximately five times more abundant than  baryonic matter in the Universe. Although DM can be in the form of more conventional
compact objects like primordial black holes~\cite{Carr:2009jm}, there is also the possibility that DM might be in the form of particles. No
Standard Model (SM) particle can play the role of DM. Therefore, in the case DM is in the form of particles, it must be related to physics beyond the SM. The simplest example of such a realization is the Weakly Interacting Massive Particle (WIMP) paradigm. In that case, WIMPs are produced in the early Universe and occasionally annihilate to SM particles. If the annihilation cross section is appropriate, annihilations are sufficient to produce the DM abundance we observe today. Such a scenario can potentially be tested. Annihilation of WIMPs in the Sun and the center of the Galaxy to SM particles
could create observable signals. Similarly, this scenario allows production of WIMPs in collider experiments as long as there is enough energy to produce them or create nuclear recoils when WIMPs scatter off  nuclei in underground detectors.

A minimal extension of the SM that could facilitate the above characteristics can be realized by adding a new  $U(1)$ gauge symmetry which breaks spontaneously providing the dark photon with a mass~\cite{Kobzarev:1966qya,Okun:1982xi}.  In this scenario, the DM particle is charged under the $U(1)$ symmetry and since it is possible to have a small kinetic mixing between the dark photon and the SM photon under very generic grounds 
\cite{Holdom:1985ag}, the dark and bright sectors are linked. As we mentioned above, a particular way of testing this scenario is by 
searching for signals of DM annihilation in the Sun. DM particles can be trapped in the Sun simply by interacting with nuclei or electrons as they cross it. Trapped DM particles may thermalize with the interior
of the Sun and sink to the center where they can meet each other and annihilate to SM particles. In particular, annihilation to neutrinos (which can easily escape from the Sun) may potentially lead to detectable signals in Earth-based detectors. The capture and annihilation process in the Earth and the Sun as well as the produced neutrino spectrum from DM annihilation has been 
studied extensively in the past~\cite{Press:1985ug,Gould:1987ir,Gould:1987ju,Silk:1985ax,Freese:1985qw,Krauss:1985aaa,Griest:1986yu}. Within the context of dark photons, similar studies have also been made  regarding indirect signals from the Sun and the Earth or direct detection~\cite{Pospelov:2007mp,Schuster:2009au,Schuster:2009fc,Meade:2009mu,An:2013yfc,An:2013yua,Fradette:2014sza,An:2014twa,Feng:2015hja,
Feng:2016ijc}.

In this paper we investigate fermionic DM that self-interacts via a light dark photon mediator. In particular, we are interested in the region of the parameter space, where DM captured by the Sun has a substantial probability of recombining to positronium-like bound states of DM and anti-DM, which we from now on call darkonium. Darkonium states can subsequently decay to either two dark photons in the case of para-darkonium (where the spin of DM and anti-DM are opposite) or to three dark photons in the case of ortho-darkonium (where the spin of DM and anti-DM are aligned). We assume that dark photons are linked to SM via kinetic mixing with the ordinary photons. At the end of the day, the spectrum of the produced dark photons in the Sun has four components: i) monochromatic dark photons of energy
$m_X$ (the mass of DM) which come from direct DM annihilation ii) monochromatic dark photons of energy $m_X-\Delta/2$ where $\Delta$ is the binding energy of the para-darkonium iii) dark photons that span energy up to $m_X-\Delta/2$ from the decay of ortho-darkonium states and iv) monochromatic photons of energy $\Delta$ produced from the recombination of DM to darkonium. We investigate under what conditions the aforementioned dark photons leave the Sun without decaying to SM particles. Those that exit the Sun intact, will decay sooner or later producing an electron-positron pair due to the kinetic mixing. Based on that, we find the precise positron spectrum that potentially could be observed by AMS-02 and we show how one can set constraints on these models. Interestingly, we find that there is parameter space where recombination dark photons could be detected earlier than the annihilation ones.

We should also mention at this point that the possibility of forming darkonium can also have important consequences in the physics of the early Universe. For instance, if DM is thermally produced by a freeze-out of DM annihilation into mediators, the effect of darkonium bound states is to delay the freeze-out until later times~\cite{vonHarling:2014kha}. Additionally, the effect of the bound states in indirect Galactic searches and on the Galactic structure has been investigated in~\cite{An:2016gad,Feng:2009mn}. The indirect detection signal in models with light dark photons in the center of the Earth and the Sun have been recently investigated in~\cite{Feng:2015hja,Feng:2016ijc} albeit neglecting DM bound states. As we will show, in the region of the parameter space where darkonium can form, the spectrum of positrons is dominated by the decay pattern of the bound states, and direct annihilation contributes a subleading part in the full signal.

The paper is structured as follows: In section \ref{Sec: DM in the Sun} we review the DM model and the necessary formalism to calculate the solar DM and darkonium populations. In section \ref{Sec: Parameter space} we demarcate the allowed parameter space in which darknonium can lead to a significant indirect detection signal. In section \ref{Sec: Kinematics of mediator decay} we review the kinematics and geometry of the positron signal from a decaying mediator emitted from the Sun. In section \ref{Sec: Results} we present the positron spectra and compare them with the flux observed by AMS-02. In section \ref{Sec: Conclusions} we make our concluding remarks.

Throughout the paper we use natural units, i.e. $\hbar = c = k_\text{B}= 1$.

\section{Dark matter in the Sun} \label{Sec: DM in the Sun}

We start by introducing a generic DM model that can accommodate the formation of darkonium bound states.
The DM model includes a Dirac fermion $X$ (and its antiparticle $\bar{X}$) which constitutes the bulk of the DM relic abundance. The DM candidate interacts through a dark $U(1)$ gauge symmetry. The associated dark photon $\phi$ interacts with the SM by kinetic mixing with the SM photon. The Lagrangian density of the model reads
\begin{equation}
\mathcal{L} = \bar{X}\left(i \gamma^\mu D_\mu -m_X\right)X - \frac{1}{4}\Phi_{\mu\nu}\Phi^{\mu\nu} + \frac{m_\phi^2}{2}\phi_\mu\phi^\mu - \frac{\epsilon}{2}\Phi_{\mu\nu}F^{\mu\nu},
\end{equation}
where $m_X$ is the DM mass, $m_\phi$ the mediator mass, $\epsilon$ the kinetic mixing parameter and $D_\mu = \partial_\mu -i g_X\phi_\mu$ is the covariant derivative. $F_{\mu\nu}$ is the electromagnetic field strength tensor and $\Phi_{\mu\nu} = \partial_\mu \phi_\nu - \partial_\nu \phi_\mu$ is the dark $U(1)$ field strength tensor. We further define $\alpha = e^2/(4\pi)$ and $\alpha_X = g_X^2/(4\pi)$ as the electromagnetic and dark fine structure constants. The kinetic mixing term can arise through integrating out a heavy particle that is charged under both the dark and electromagnetic gauge groups \cite{Holdom:1985ag, Collie:1998ty}, leading naturally to a small $\epsilon$. For the purposes of this paper we will treat $\epsilon$ as a free, yet small, parameter.
 
 The above generic model allows for self-interactions of DM of Yukawa type since 
  the dark photon has a mass (either through a Higgs- or Stueckelberg-mechanism) of the form
 \begin{equation}
 V=\pm \alpha_X \frac{e^{-m_\phi r}}{r},
 \end{equation}
 $r$ being the distance between two DM particles.
The potential is repulsive ($+$ sign) for $XX$ or $\bar{X}\bar{X}$ interactions, while $X\bar{X}$ interactions are attractive ($-$ sign) and can lead to formation of bound states.  In fact
DM self-interactions are welcome since they can solve a range of problems arising in the collisionless cold DM paradigm such as  
 \emph{the too big to fail problem}~\cite{BoylanKolchin:2011de}, \emph{the missing satellites problem}~\cite{Klypin:1999uc,Moore:1999nt,Kauffmann:1993gv,Liu:2010tn,Tollerud:2011wt,Strigari:2011ps}
 and \emph{the core-cusp problem}~\cite{Moore:1994yx, Flores:1994gz, Navarro:1996gj}. Numerical simulations  suggest that generally speaking DM self-interactions alleviate the aforementioned problems  within approximately the range of $\sigma/m_X$ ($\sigma$ being the DM-DM cross section) $0.1-1\,\text{cm}^2/\text{g}$~\cite{Rocha:2012jg,Peter:2012jh}. 
  
The model has four free parameters: $\epsilon$, $m_\phi$, $m_\chi$ and $\alpha_X$. In this paper, we assume that the DM relic abundance $\Omega_X \simeq 0.23$ is produced though the thermal freeze-out of $X+\bar{X} \to \phi +\phi$, i.e. the DM density is fixed once the annihilation rate drops below the Hubble expansion rate, $\Gamma (X + \bar{X}\to \phi +  \phi) \lesssim H$. We use this criterion to fix the value of $\alpha_X$ as a function of $m_X$. The relation between $\alpha_X$ and $m_X$ has been calculated in detail in \cite{vonHarling:2014kha} including the effects of both Sommerfeld enhancement and darkonium recombination\footnote{We note that the recombination cross sections used in \cite{vonHarling:2014kha} differs from that used in \cite{An:2016gad}. The coupling $\alpha_X$ fixed by the DM relic abundance in \cite{vonHarling:2014kha}  differs slightly from the result of direct annihilation with Sommerfeld enhancement, while \cite{An:2016gad} claims that the effect of recombination on freeze-out is negligible all together. For the purposes of removing $\alpha_X$ as a free parameter we adopt the coupling found by \cite{vonHarling:2014kha}.}.

The decay rate of the dark photon into SM fermions is given by \cite{Feng:2016ijc}
\begin{equation}
	\Gamma(\phi \to f\bar{f}) = \frac{\epsilon^2 q_f^2\alpha(m_\phi^2 + 2m_f^2)}{3m_\phi^2} \sqrt{1-\frac{4m_f^2}{m_\phi^2}},
\end{equation}
where $q_f$ is the electric charge of the fermion (in units of $e$) and $m_f$ is the fermion mass. The decay length into positron electron pairs is
\begin{equation}
	L = \text{Br}(\phi \to e^+e^-) \left( \frac{1.1 \cdot 10^{-9}}{\epsilon}\right)^2 \left( \frac{m_X/m_\phi}{1000}\right) \left(\frac{100 \text{MeV}}{m_\phi} \right) R_\odot,
	 \label{Eq: decay length}
\end{equation} 
where $R_\odot = 7.0 \cdot 10^{8}$ m is the radius of the Sun, and the branching ratio to electron-positron pairs $\text{Br}(\phi \to e^+e^-) \simeq 1$, if the dark photon cannot decay to heavier particles, e.g. to muons $m_\phi < 2m_\mu = 211$ MeV. The branching ratios for heavier dark photons has been calculated in \cite{Buschmann:2015awa}.

\subsection{Solar capture} \label{Sec: Solar Capture}
We now proceed to calculate the number of DM particles captured in the Sun's gravitational field after scattering on nuclei. We follow the procedure described in \cite{Gould:1987ir,Feng:2016ijc}. The capture rate of DM in the Sun for a particular nuclear species $N$ is given by 
\begin{equation}
C_\text{cap}^N = n_X^\text{(loc)} \int d^3 \vec{r}\;d^3\vec{w}\;  n_N(r) w f_\odot(w, r) \int dE_R \frac{d \sigma_N}{dE_R},
\end{equation}
where $n_X^\text{(loc)}  = (\rho_\text{DM}/2)/m_X= (0.2$ GeV/cm$^3)/m_X$ is the local number density of $X$ particles, $n_N(r)$ is the number density of the nuclear species $N$ as a function of the distance to the center of the Sun, $w$ is the DM-nucleus relative velocity, $f_\odot(w, r)$ is the DM velocity distribution in the rest frame of the Sun, $E_R$ is the recoil energy and $d\sigma_N/dE_R$ is the elastic differential scattering cross section of DM on the nucleus $N$. By energy conservation $w$ is related to the DM velocity asymptotically far from the Sun $u$ by
\begin{equation}
w^2 = u^2+ \frac{2GM(r)}{r}. \label{Eq: relative velocity w}
\end{equation}
The velocity distribution in the galactic rest frame is taken to be \cite{Baratella:2013fya}
\begin{equation}
f(u) = N \left[\exp \left(\frac{v_\text{gal}^2-u^2}{ku_0^2} \right)-1 \right]^k \Theta(v_\text{gal}-u),
\end{equation}
where we set the galactic escape velocity to be $v_\text{gal} = 550$ km/s and $u_0 = 245$ km/s. The parameter $k$ lies in the interval $1.5 \leq k \leq 3.5$. The normalization $N$ is chosen such that $\int d^3 \mathbf{u} f(u) = 1$. Liouville theorem dictates that $f_\odot(w, r)=f(\vec{u}+\vec{v}_{\odot})$ ($\vec{v}_{\odot}$ being the velocity of the Sun with respect to the halo).
In the rest frame of the Sun the boosted velocity distribution is
\begin{equation}
\bar{f}(u) = \frac{1}{2} \int_{-1}^1 dc f\left(\sqrt{u^2 + u_\odot^2 + 2 u u_\odot c}\right).
\label{Eq: velocity distribution before integration}
\end{equation}
The final capture rate  has only  a mild dependence on $k$. We  choose $k=2$, so we can integrate Eq.~\ref{Eq: velocity distribution before integration} analytically:
	\begin{equation}
	 \bar{f}(u)= \frac{N}{4 u u_\odot}
    \begin{dcases*} 
        e^{-\tfrac{(u+u_\odot)^2}{u_0^2}} \left[ 4 e^{\tfrac{(u+u_\odot)^2}{u_0^2}}u u_\odot - 4e^{\tfrac{(u+u_\odot)^2+v_\text{gal}^2}{2u_0^2}}\left(e^{\tfrac{2uu_\odot}{u_0^2}} -1\right)u_0^2 + e^{\tfrac{v_\text{gal}^2}{u_0^2}} \left(e^{\tfrac{4uu_\odot}{u_0^2}}-1 \right)u_0^2\right] \\
         \hspace{10cm} \text{if} \; \; u\leq v_\text{gal}-u_\odot \\ 
      e^{-\tfrac{u^2+u_\odot^2}{u_0^2}}\left[ e^{\tfrac{2 u u_\odot + v_\text{gal}^2}{u_0^2}}u_0^2 - 4 e^{\tfrac{(u+u_\odot)^2+ v_\text{gal}^2}{2u_0^2}}u_0^2 + e^{\tfrac{u^2+u_\odot^2}{u_0^2}}\left(3 u_0^2 - (u-u_\odot)^2 + v_\text{gal}^2\right) \right] \\
        \hspace{8.3cm} \text{if} \; \; v_\text{gal}-u_\odot < u\leq v_\text{gal}+u_\odot
    \end{dcases*}.
\end{equation}
The differential cross section for a non-relativistic elastic scattering is \cite{Feng:2015hja, Feng:2016ijc}
\begin{equation}
\frac{d\sigma_N}{dE_R} = \frac{8\pi \epsilon^2 \alpha_X \alpha Z_N^2 m_N}{w^2 (2m_N E_R + m_\phi^2)^2} \left| F_N\right|^2,
\end{equation}
where $m_N$ is the mass of the target nucleus, $Z_N$ is the number of protons and the Helm form factor is given by $|F_N|^2 = \exp(-E_R/E_N)$ with $E_N = 0.114\,A_N^{-5/3} \, \text{GeV} $. The recoil energy must be at least $E_\text{min} = m_X u^2/2$ for DM to be captured in the Sun's gravitational potential in a single scattering. The maximum recoil energy in an elastic scattering is $E_\text{max} = 2 \mu_N^2 w^2/m_N$, where $\mu_N = m_X m_N/(m_X+m_N)$ is the DM-nucleus reduced mass. With these integration limits the capture rate can be expressed as \cite{Feng:2015hja}
\begin{equation}
C_\text{cap}^N = \frac{32 \pi^3 \epsilon^2 \alpha_X \alpha n_X Z_N^2}{m_N E_N}\exp\left(\frac{m_\phi^2}{2m_N E_N} \right) c_\text{cap}^N,
\label{Eq: capture rate}
\end{equation}
where $c_\text{cap}^N$ is the integral
\begin{equation}
c_\text{cap}^N = \int_0^{R_\odot} dr\, r^2 n_N(r)\int_0^\infty du\, u\bar{f}(u)  \times \Theta(\Delta x_N) \left[\frac{e^{-x_N}}{x_N} + \text{Ei}(-x_N)\right]_{x_N^\text{max}}^{x_N^\text{min}},
\end{equation}
with $x_N = (2m_N E_R + m_\phi^2)/(2m_N E_N)$ and $\text{Ei}(z) = -\int_{-z}^\infty dt\, t^{-1} \exp(-t)$. The total capture rate is given by $C_\text{cap} = \sum_N C_\text{cap}^N$, where the density profiles of relevant nuclei are taken from the solar composition model AGSS09\footnote{The solar composition model is publicly available at \url{http://wwwmpa.mpa-garching.mpg.de/~aldos/solar_main.html}.}~\cite{Serenelli:2009yc,Serenelli:2009ww}. We make the simplifying approximation that the composition of the Sun has been well-described by this model throughout its lifetime.

\subsection{Annihilation of free dark matter}
When DM particles interact through a light mediator, and the relative velocity between annihilating DM particles is low, the cross section is significantly enhanced compared to the tree-level annihilation cross section~\cite{Cassel:2009wt,Slatyer:2009vg,Feng:2010zp}. This effect is parametrised by the Sommerfeld enhancement factor $S$
\begin{equation}
\langle \sigma_\text{ann} v\rangle = S\; \langle\sigma_\text{ann}v \rangle_0
\end{equation}
where $\sigma_\text{ann}\equiv\sigma(X + \bar{X} \to \phi+\phi)$ and
\begin{equation}
\langle\sigma_\text{ann}v \rangle_0=\frac{\pi \alpha_X^2}{m_X^2} \frac{(1-m_\phi^2/m_X^2)^{3/2}}{(1-m_\phi^2/(2m_X^2))^2},
\end{equation}
is calculated in the Born regime. The Sommerfeld enhancement factor for the $s-$wave process can be found using Hulth\'en's potential to approximate the Yukawa potential, yielding
\begin{equation}
S_s = \frac{\pi}{a} \frac{\sinh(2\pi ac)}{\cosh(2\pi ac)-\cos(2\pi \sqrt{c-a^2c^2})},
\end{equation}
where $a= v/(2\alpha_X)$ and $c = 6\alpha_X m_X /(\pi^2 m_\phi)$. We take the full Sommerfeld-enhancement factor to be $S \approx \langle S_s \rangle$, where $\langle \cdot \rangle = \int d^3v e^{-\tfrac{1}{2}v^2/v_0^2}/(2\pi v_0^2)^{3/2}$ is the thermal average, with $v_0$ the typical relative velocity. When the annihilating DM particles are non-relativistic, the dark photons will have energy $E_\phi \simeq m_X$. The distribution of dark photons as a function of energy when DM particles undergo a direct annihilation process is therefore a delta function,
\begin{equation}
\label{Nann}
	\frac{dN_\phi^{(\text{ann})}}{dE_\phi} = 2\delta(E_\phi - m_X).
\end{equation}

\subsection{Recombination}
Darkonium bound states ($D$) in this model can form by emission of an on-shell dark photon. At low energies the cross section for forming darkonium may be larger than direct annihilation, i.e. $\sigma_\text{rec}>\sigma_\text{ann}$ where $\sigma_\text{rec}\equiv\sigma(X + \bar{X} \to D+\phi)$. This suggests that recombination can have a big impact on DM freeze-out (as noted in \cite{vonHarling:2014kha}) and on indirect detection. If the binding energy of darkonium is less than the mass of the dark photon, the radiated recombination photon must be virtual. The cross section is in this scenario suppressed by a factor $\epsilon^2$, and darkonium formation becomes negligible. 
In the Coulomb limit $m_\phi\to 0$, the binding energy of the $n$th excited state is $\Delta_n=\alpha_X^2m_X/(4 n^2)$ (henceforth we drop the index on the ground state binding energy, i.e. $\Delta \equiv \Delta_1$). The recombination cross section to the $n$th state has been estimated in \cite{BetheSalpeter_QM, Lifshitz_RelativisticQM}. 
Having a non-zero mediator mass means that there is a highest state which can be populated, $n_\text{max}=\alpha_X\sqrt{m_X/(4m_\phi)}$. Beyond $n_\text{max}$ the binding energy is too low to emit an on-shell dark photon. Although highly excited states decay slowly to dark photons,  they can make a transition quickly to lower states, if the difference in binding energy allows for emission of an on-shell dark photon.
%
The highest excited state $\tilde{n}$, where  transitions from $\tilde{n}$ to $\tilde{n}-1$ can take place by emission of an on-shell $\phi$ is $\tilde{n} \approx (\alpha_X^2m_X/(2m_\phi))^{1/3}$ when $\tilde{n}\gg 1$. Thus the $\tilde{n} \approx (2n_\text{max}^2)^{1/3}$ lowest lying states can quickly cascade to the $n=1$ state. Since the transition is quick, we will treat all darkonium decays of states below $\tilde{n}$ as decays of the ground state. We note that the $2S$-state can also decay \cite{Alonso:2016faw}, but we expect that ignoring this introduces only a small error in our estimate. In the case of a non-zero mediator mass, we do not have an analytic expression for the wavefunctions of neither the discrete nor the continuous spectrum. Instead we obtain the recombination cross section by partial wave expansion and solving the Schr\"odinger equation approximating energy levels and out-going wave functions to be those of the hydrogen atom. We detail in appendix \ref{App: Schrodinger Eq} our numerical procedure for obtaining the recombination cross section.  The procedure uses as a first step the same approach as \cite{Tulin:2013teo} for calculating the scattering cross section.

The authors of \cite{An:2016gad} recently found that the recombination cross section of the present DM model is much larger than that of direct DM annihilation, when the relative velocity of an $X\bar{X}$ pair is in the interval $2m_\phi/m_X < v < 2\sqrt{m_\phi/m_X}$, where $v$ is the relative velocity of the $X \bar{X}$ pair. In this region the two are approximately related by\footnote{This relation is most correct in the limit where many darkonium states can be populated. If only a few states can be populated by emission of an on-shell dark photon, the relation breaks down.}
\begin{equation}
	\langle\sigma_\text{rec} v\rangle \approx \frac{64}{3\sqrt{3}\pi} \ln \left(\frac{\alpha_X}{2}\sqrt{\frac{m_X}{m_\phi}} \right)\langle\sigma_\text{ann} v\rangle.
	\label{Eq: Recombination rate when it's large}
\end{equation}
This cross section accounts for all darkonium states up to $n_\text{max}$. If the relative velocity $v\ll 2m_\phi/m_X$ the recombination rate becomes suppressed with respect to the annihilation rate. 
We emphasise that the indirect detection signal from recombination photons can dominate over the annihilation signal in particular energy bins, even if $\sigma_\text{rec}< \sigma_\text{ann}$. This is related to the fact that the positron spectrum comes in flat box-shapes, as we will discuss in detail in section \ref{Sec: Kinematics of mediator decay}. The width of the box spectrum is determined by the mediator energy. In the case of recombination, the mediator energy is the ground state binding energy $\Delta$, whereas in direct annihilation  is $m_X$. The box from direct annihilation spans over a much larger energy range than recombination photons. Thus even if $\sigma_\text{rec}< \sigma_\text{ann}$, recombination photons although fewer than annihilation ones, span over a much smaller energy range and therefore can dominate the low energy bins.


Neglecting hyperfine splitting, ground state ($1S$) darkonium can be in the spin zero singlet state para-darkonium ($p$), or the spin one triplet state ortho-darkonium ($o$). At leading order the $p$-darkonium state decays back-to-back (in the rest frame) into two dark photons. The photons from the $p$-darkonium state are thus distributed in energy as
\begin{equation}
	\label{Ndp}
	\frac{dN_\phi^{(p)}}{dE_\phi} = 2\delta(E-m_X-\Delta/2).
\end{equation}
The $o$-darkonium state on the other hand decays at leading order into three dark photons. As a consequence the spectrum of dark photons is more complicated. The photon spectrum is similar to that from decaying ortho-positronium which was calculated for the first time in 1949 by Ore and Powell \cite{Ore:1949te}. According to the Ore-Powell spectrum, the most likely decay (in the rest frame) is into two nearly back-to-back photons along with a single soft photon. In \cite{An:2016gad} and \cite{An:2015pva} An et al. calculated the distribution of bound state decays into three dark photons taking the mass of the dark photon into account. They found the following formula
\begin{equation}
	\frac{dN_\phi^{(o)}}{dE_\phi} = \frac{9}{4(\pi^2-9)y^2}\frac{  y(8-3y) + \frac{(y-1)}{y-2}(y^2-6y+16)\log(1-y)}{m_X - \Delta/2 - m_\phi - 3m_\phi^2/(4m_X-2\Delta)},
	\label{Eq: Ore-Powell spectrum}
\end{equation}
where $y=E_\phi/(m_X-\Delta/2)$ is the dark photon energy in units of the darkonium mass, which runs from $y_\text{min} = m_\phi/(m_X -\Delta/2)$ to $y_\text{max} = 1-3m_\phi^2/(16m_X -8\Delta)^2$.

Dark photons emitted in the recombination process will have energy equal to the binding energy of darkonium. Again we neglect the kinetic energy of the recombining DM. Every time a pair of $X \bar{X}$ recombines we obtain one recombination photon. We approximate the distribution in energy to be a delta function centered at the ground state energy
\begin{equation}
	\label{Nrec}
	\frac{dN_\phi^{(\text{rec})}}{dE_\phi} \simeq \delta(E_\phi - \Delta).
\end{equation}
This approximation is best in the limit, where only few states can be populated by emission of an on-shell dark photon.
\subsection{Boltzmann equations}

Here we combine the rates described in the previous subsection to calculate the DM population in the Sun. We assume that DM particles  thermalize in a time scale much shorter than the other time scales we are going to consider. A posteriori this approximation is justified in the parameter space we are interested in. DM in the Sun can be either free or bound in darkonium states. The abundances of each component are described by a system of Boltzmann equations given by
\begin{subequations}
\begin{align}
\frac{dN_X}{dt} &= C_\text{cap}-(C_\text{ann}+C_\text{rec})N_X^2, \label{Eq:Boltzmann eqs a} \\
\frac{dN_{o}}{dt} &= \frac{3}{4}C_\text{rec}N_X^2  - C_\text{o}N_o, \label{Eq:Boltzmann eqs b} \\
\frac{dN_{p}}{dt} &= \frac{1}{4}C_\text{rec}N_X^2  - C_\text{p}N_p. 
\label{Eq:Boltzmann eqs c}
\end{align}
\end{subequations}
Here $N_X$ is the number of free DM particles $X$. We omit the equation for anti-DM, since DM is symmetric and $N_X=N_{\bar{X}}$ at all times. The number of darkonium in $o$- and $p$-states are given by $N_{\text{o/p}}$, $C_\text{cap}$ is the rate at which the Sun captures $X$-particles from the DM halo (not counting anti-DM) as described in section \ref{Sec: Solar Capture}, and $C_{o/p}$ are the decay rates of the $o$- and $p$-states, which are given by \cite{Stroscio:1975fa}
\begin{align}
	C_o &= \frac{2(\pi^2-9)}{9\pi} \alpha_X^6 m_X \simeq 0.06 \alpha_X^6 m_X,\notag \\
	C_p &=  \frac{\alpha_X^5}{2} m_X.
\end{align}
We consider DM heavier than $\sim$ 4 GeV and therefore  particle evaporation is negligible~\cite{Gaisser:1986ha,Griest:1986yu}. The rates of annihilation and recombination are obtained by integrating over the DM number density inside the Sun, i.e. $C_\text{ann} = N_X^{-2}\int d^3 \mathbf{x} \; n_X^2(r) \langle \sigma_\text{ann} v \rangle$ and similar for $C_\text{rec}$. The distribution of DM inside the Sun is given by
\begin{equation}
n_X(r) = n_0\exp\left(-\frac{r^2}{r_\text{th}^2} \right),
\end{equation}
where $n_0$ is the central number density of $X$-particles and $r_\text{th}$ is the thermal radius, which by the virial theorem is\begin{equation}
r_\text{th} = \sqrt{\frac{3T_\odot}{2\pi G \rho_\odot m_X}}.
\end{equation}
$T_\odot\approx 1.55 \cdot 10^7$ K and $\rho_\odot \approx 151$ g/cm$^3$ are the temperature and density in the center of the Sun, respectively. Taking $N_X = \int d^3 \mathbf{x}\; n_X(r)$ the annihilation and recombination rates become
\begin{align}
	C_\text{ann} &=\frac{ \langle \sigma_\text{ann} v \rangle}{(2\pi)^{3/2}r_\text{th}^3}, \notag \\
	C_\text{rec} &=\frac{ \langle \sigma_\text{rec} v \rangle}{(2\pi)^{3/2}r_\text{th}^3}.
\end{align}
 The factors 1/4 and 3/4 in equations \ref{Eq:Boltzmann eqs b} and \ref{Eq:Boltzmann eqs c} denote the different multiplicities of the ortho- and para- states i.e. the ortho-darkonium has three spin states, while the para only  one. In Eqs.~\ref{Eq:Boltzmann eqs a} - \ref{Eq:Boltzmann eqs c} we have neglected a number of subdominal effects: ionization of darkonium, populating higher excited darkonium states and DM self-capture. In appendix \ref{App: Verification of approximations} we verify that these effects can be neglected for the parameter space we investigate.
The Boltzmann equations Eqs.~\ref{Eq:Boltzmann eqs a} - \ref{Eq:Boltzmann eqs c} admit the analytical solutions 
\begin{subequations}
\begin{align}
N_X(t) &= \sqrt{\frac{C_\text{cap}}{C_\text{ann}+C_\text{rec}}} \tanh  \frac{t}{\tau_X},
\label{Eq: Boltzmann solution a}\\
N_i(t) &=\frac{e^{-C_i t}}{2C_i(2+C_i\tau_X)} \left\{ 2e^{\left(C_i+\tfrac{2}{\tau_X}\right)t}C_i^2 q_i\tau_X^2 {}_2F_1\left(1,\tfrac{2+C_i\tau_X}{2},\tfrac{4+C_i\tau_X}{2},-e^{\tfrac{2t}{\tau_X}}\right) +\right. \notag \\
& \qquad q_i(2+C_i\tau_X) \left[ 2\left(e^{C_it} - C_i \tau_X -1\right)  + C_i^2\tau_X^2\left(  \psi_0(\tfrac{2+C_i\tau_X}{4})-\psi_0(\tfrac{C_i\tau_X}{4})\right) \right.-\notag \\
&\qquad \left.\left. 2 C_i \tau_X e^{C_it} \left( {}_2F_1\left(1,\tfrac{C_i\tau_X}{2},\tfrac{2+C_i\tau_X}{2},-e^{\tfrac{2t}{\tau_X}}\right) + \tanh \frac{t}{\tau_X}\right) \right]\right\},
\label{Eq: Boltzmann solution b}
\end{align}
\end{subequations}
where $N_i\; (C_i)$ refer to either $N_o\; (C_o)$ or $N_p\; (C_p)$, the $q_i$s are given by $q_o = 3 C_\text{rec} C_\text{cap}/(4(C_\text{rec}+ C_\text{ann}))$ and $q_p =q_o/3$, $\psi_0(x) = \Gamma'(x)/\Gamma(x)$ is the digamma function and ${}_nF_m$ are the hypergeometric functions. The characteristic time $\tau_X$ before $N_X$ reaches the steady state is given by
\begin{equation}
\tau_X = \frac{1}{\sqrt{C_\text{cap}(C_\text{ann}+C_\text{rec})}},
\label{Eq:steady state time}
\end{equation}
whereas the characteristic steady state time scale for darkonium is
\begin{equation}
	\tau_{o/p} = \max\left\{\tau_X,\; \frac{1}{C_{o/p}} \right\}.
	\label{Eq: Darkonium steady state time}
\end{equation}

\begin{figure}[h!]
	\includegraphics[width=.5\textwidth]{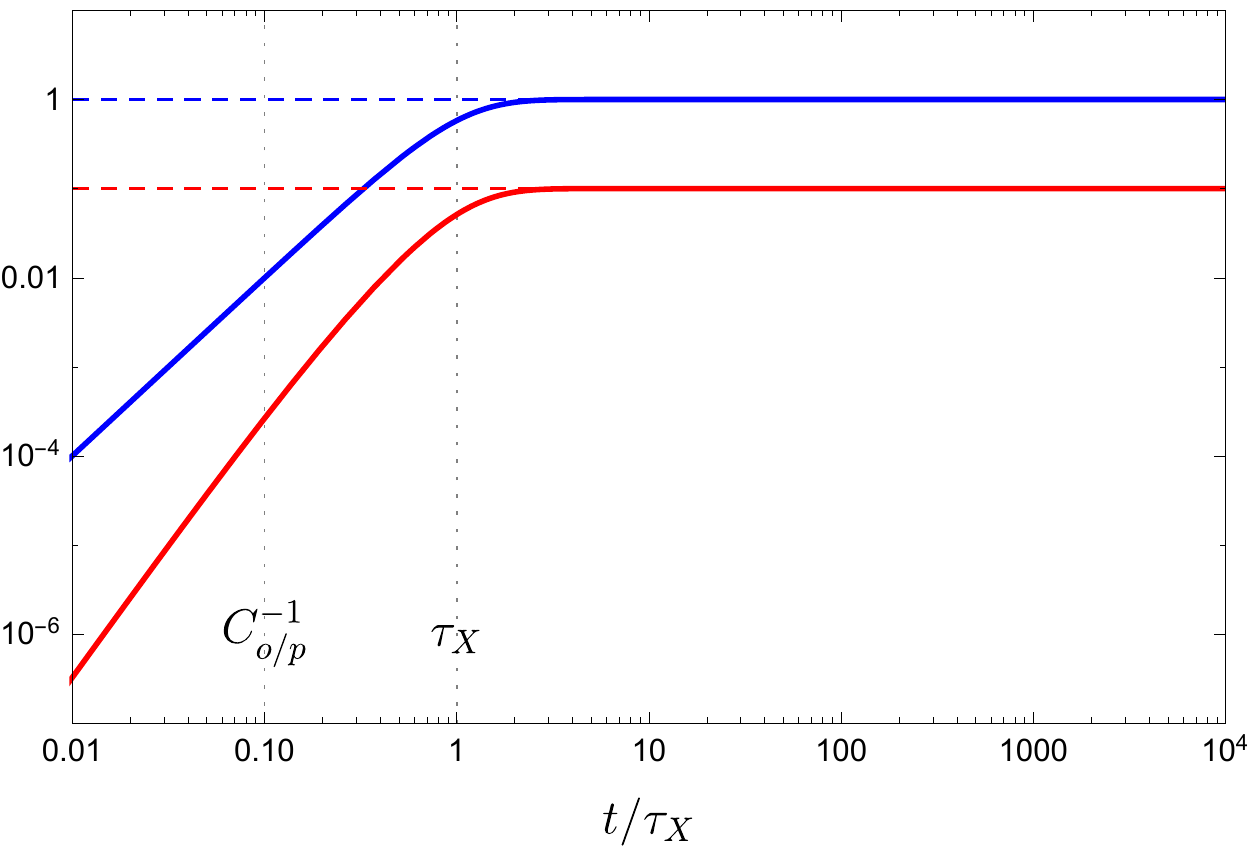}
		\includegraphics[width=.5\textwidth]{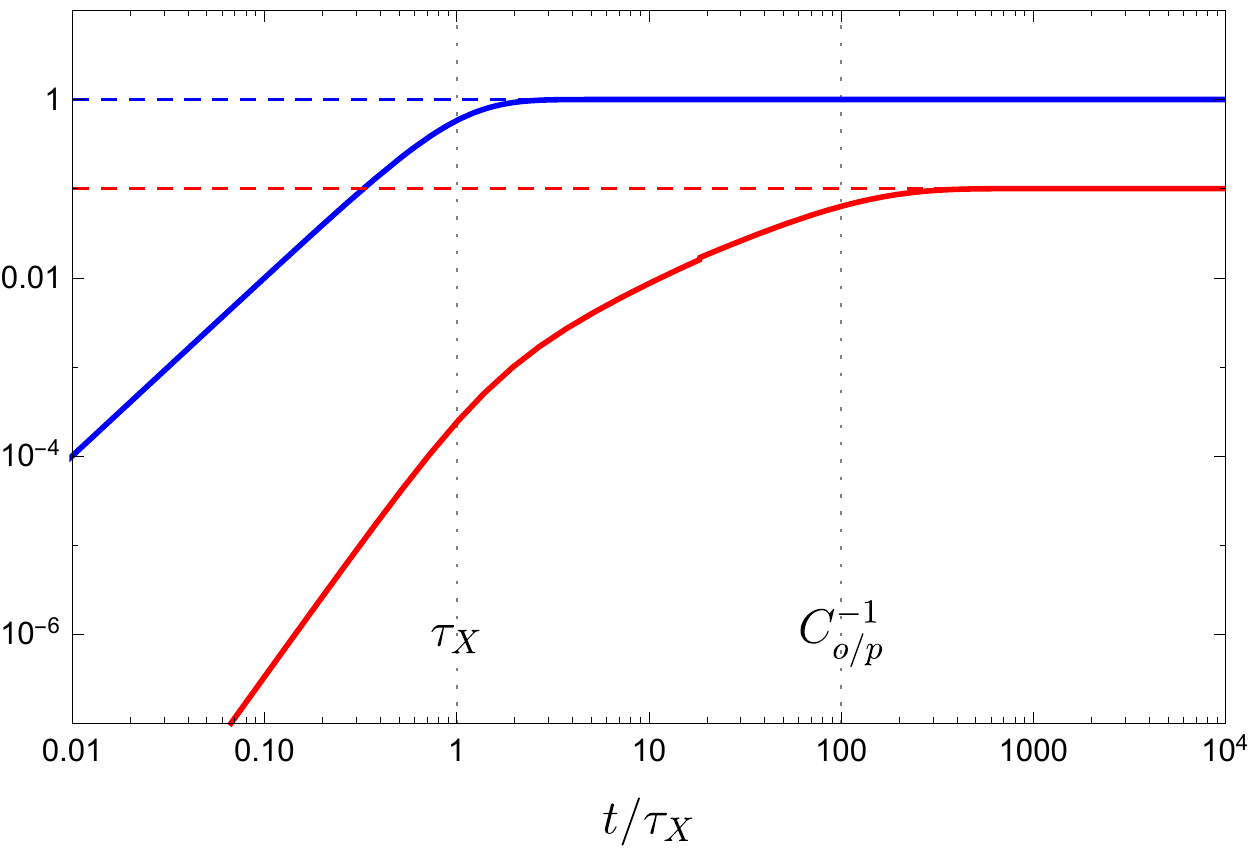}
		\caption{This figure shows the solar population of free DM (blue) and darkonium (red) as a function of time illustrating the relevant time scales involved. We have arbitrarily normalised  $N_X^\text{ss} = 1$ (dashed blue line) and $N_i^\text{ss} = 1/10$ (dashed red line). In the left figure we have chosen $C_{o/p}^{-1} = \tau_X/10$, and in the right figure we set $C_{o/p}^{-1} = 100\,\tau_X$. The figures reflect the fact that the time scale for steady state for the darkonium population is given by Eq.~\ref{Eq: Darkonium steady state time}.}
		\label{Fig: Boltzmann equations.}
\end{figure}

For all the parameters we will consider $1/C_{o/p} \ll \tau_X$ and therefore $\tau_{o/p} = \tau_X$.
Once the DM  has reached the steady state,  the populations are
\begin{align}
N_X^\text{ss} &= \sqrt{\frac{C_\text{cap}}{C_\text{ann}+C_\text{rec}}}, \notag \\
N_{o/p}^\text{ss} &= \frac{q_{o/p}}{C_{o/p}}. \label{Eq: Steady states}
\end{align}
The free DM and darkonium populations in the Sun described by Eqs.~\ref{Eq: Boltzmann solution a} and \ref{Eq: Boltzmann solution b} are shown in Fig.~\ref{Fig: Boltzmann equations.}. The behaviour of the darkonium build-up has two qualitatively different behaviours depending on whether $\tau_X$ is smaller or larger than $C_{o/p}^{-1}$.  We require darkonium to have reached the steady state population within the lifetime of the Sun, such that the dark photon emission is not suppressed by a too slow build-up of DM, i.e. we require $\tau_{o/p}<\tau_\odot \approx 4.6$ Gyr. Since the inequality $\tau_X > C_{o/p}^{-1}$ is true for all parameters we will consider, it will suffice to find the region where $\tau_X< \tau_\odot$.


\section{Parameter space} \label{Sec: Parameter space}

In this section we map the allowed and interesting region of the parameter space. The parameter space allowed by observations, which simultaneously permits bound state formation, is quite complex. It is beyond the scope of this paper to make a detailed analysis and scan over all possible parameter sets. Instead we list  the most important constraints within the interesting region of darkonium formation. The available regions in parameter space is demarcated by the requirements listed below. Bullets 1-5 are regions of interest (\emph{not} constraints), while bullets 6-8 contain observational constraints.

\begin{enumerate}
\item \textbf{Bound state formation:} Darkonium bound states can in principle exist if $1/m_\phi$ is longer than the Bohr length, $a_0 = 2/(\alpha_X m_X)$. However, a stronger requirement arises by demanding the darknonium to be able to form in the first place by emission of an on-shell dark photon. This condition can be written as $m_\phi< \Delta=\alpha_X^2 m_X/4$.
\item \textbf{Electron-positron decay:} We require that the dark photon can decay into an electron-positron pair, i.e. $m_\phi>2m_e$. Within the model we consider, this inequality is always fulfilled if the observational constraints 6-8 are obeyed. We would also like the branching into electron-positrons to be close to one, and therefore keep $m_\phi < 2m_\mu$.
\item \textbf{Scattering unitarity:} To keep our model self-consistent we need to impose unitarity of DM self-scattering. This limits the coupling from above to be $\alpha_X\lesssim0.54$. Correspondingly, by assuming a thermal relic DM abundance limits the DM mass below $m_X \lesssim 139$ TeV \cite{vonHarling:2014kha}.
\item \textbf{Efficient recombination:} It is useful to note the Coulombic regime where the mediator mass is smaller than the transferred momentum $m_\phi <\mu v$ but larger than the kinetic energy of the scattering, $m_\phi>\mu v^2/2$. If the lower limit is respected it is a good approximation to use Coulombic energy levels~\cite{An:2016gad}. The region can be written as $2m_\phi/m_X < v < 2 \sqrt{m_X/m_\phi}$. Within this range the recombination cross section is much larger than the direct annihilation cross section and approximately given by the analytic expression of Eq.~\ref{Eq: Recombination rate when it's large}. 
In the Sun the velocity of a thermalized DM particle is 
\begin{equation}
v_\text{th}=\sqrt{\frac{2T_\odot}{m_X}} \approx 5.1\times 10^{-5}\sqrt{\frac{\text{TeV}}{m_X}}.
\end{equation}
Taking the relative velocity of colliding DM particles to be $v = \sqrt{2} v_\text{th}$, the lower limit translates to the following inequality on the mediator mass
\begin{equation}
m_\phi < \frac{v_\text{th} m_X}{\sqrt{2}}= 36 \text{MeV} \sqrt{\frac{m_X}{\text{TeV}}}.
\end{equation}
We stress that satisfying this inequality only ensures that the analytic expression in Eq.~\ref{Eq: Recombination rate when it's large} is well described when $n_\text{max}\gg 1$. For somewhat heavier mediators the recombination rate may still be larger than the direct annihilation rate, although Eq.~\ref{Eq: Recombination rate when it's large} breaks down. As already discussed, even if the recombination rate is far smaller, the indirect signal from recombination photons may be important nonetheless.

\item \textbf{Solar steady state:} For the solar signal to be maximal, the Sun's age must be longer than the time it takes to reach the darkonium steady state population, i.e. $\tau_{o/p} < \tau_\odot \approx 4.6$ Gyr, where $\tau_{o/p} = \tau_X$ for the parameters we will consider, and is thus given by Eq.~\ref{Eq:steady state time}.

\item \textbf{Self-interaction constraints:} $N$-body simulations suggests that DM self-interactions may flatten the core of dwarf spheroidal galaxies,  alleviating tension with observations. Self-interactions stronger than roughly $0.1-10 \text{cm}^2/\text{g}$ will however reduce the central densities too much. Furthermore, at higher velocity dispersions, the ellipticity of the Milky Way is threatened if the cross section is stronger than roughly $0.1-1 \text{cm}^2/\text{g}$ \cite{Tulin:2013teo}. The scattering cross section when $m_X v/\alpha_X \gg 1$ is well described by the classical momentum transfer cross sections defined by $\int d\Omega (1-\cos\theta) d\sigma/d\Omega$ \cite{Khrapak1, Khrapak:2014xqa}
\begin{equation}
\sigma_\text{att} = 
\begin{dcases}
    \tfrac{4\pi}{m_\phi^2}\beta^2\log\left(1+\beta^{-1} \right) &  \beta \lesssim 10^{-1} \\
    \tfrac{8\pi}{m_\phi^2}\beta^2\left(1+1.5\beta^{1.65} \right)^{-1} &  10^{-1} \lesssim \beta \lesssim 10^{3} \\
       \tfrac{\pi}{m_\phi^2}\left(1+ \log\beta - \tfrac{1}{2\log\beta} \right)^{2} &  \beta \gtrsim  10^{3}
\end{dcases},
\end{equation}
in the case of attractive interactions and
\begin{equation}
\sigma_\text{rep} = 
\begin{dcases}
    \tfrac{2\pi}{m_\phi^2}\beta^2\log\left(1+\beta^{-1} \right) &  \beta \lesssim 1 \\
    \tfrac{\pi}{m_\phi^2}\left(\log 2\beta-\log\log 2\beta \right)^{2} &    \beta \gtrsim 1
\end{dcases},
\end{equation}
for repulsive ones, where $\beta = 2\alpha_X m_\phi/(v m_X)$. The typical $v$ for dwarf galaxies is of the order of $v_0\sim 10$ km/s, while galactic velocity dispersions are significantly larger at the order of $v_0\sim 200$ km/s in the case of the Milky Way. Since symmetric DM with a vector mediator interacts attractively between $X\bar{X}$ pairs and repulsively between $XX$ or $\bar X \bar X$, the total cross section per mass is $\langle \sigma\rangle/m_X = \langle \sigma_\text{att} + \sigma_\text{rep}\rangle/(2m_X)$. Here $\langle \cdot \rangle = \int d^3v e^{-\tfrac{1}{2}v^2/v_0^2}/(2\pi v_0^2)^{3/2}$ is the velocity average. The DM self-interactions are mapped in Fig.~\ref{Fig: Zurek Figure}.

\begin{figure}[h!]
\centering
\includegraphics[width=.5\textwidth]{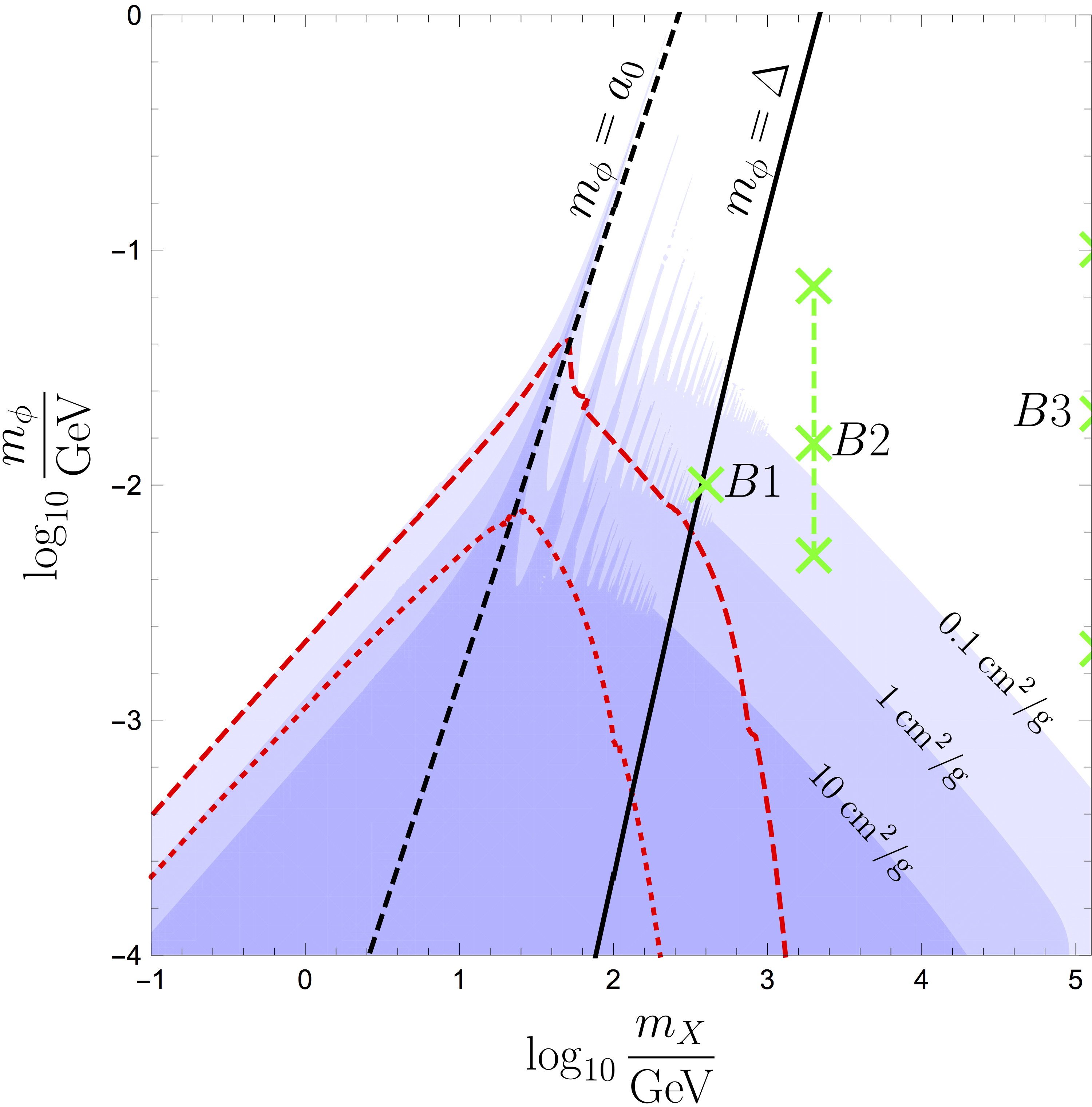}
\caption{This figure shows the $m_\phi$ vs $m_X$ parameter space. The dark coupling $\alpha_X$ is fixed such that $\Omega_X \simeq 0.23$. The blue shaded contours map DM self-interactions. From darkest to lightest blue the regions are $\langle \sigma\rangle/m_X>0.1$, $1$ and $10$ cm$^2/$g, with $\langle \sigma\rangle$ evaluated at dwarf galaxy velocity dispersion $v_0 \sim 10$ km/s. The red dotted (dashed) contours correspond to $\langle \sigma\rangle/m_X = 1 $ cm$^2/$g ($\langle \sigma\rangle/m_X = 0.1 $ cm$^2/$g) at Milky Way velocity dispersion $v_0\sim 200$ km/s. To the left of the black dashed line bound states cannot exist. To the left of the black solid line, bound states cannot recombine by emission of an on-shell dark photon. The green crosses correspond to our benchmark values.}
\label{Fig: Zurek Figure}
\end{figure}

\item \textbf{Direct detection:} Direct detection places an upper bound on $\epsilon$ for a particular choice of $m_X$ and $m_\phi$. Currently the strongest limits are placed by the LUX-experiment's 2013 results \cite{Akerib:2013tjd}. Following the procedure of \cite{DelNobile:2015uua}, we find the exclusion contours at 90\% confidence level in the $\epsilon$ versus $m_\phi$ parameter space summarized in Fig.~\ref{Fig:DD+SS constraints}. For heavy mediator masses, $\epsilon$ is less constrained. Notice, that the bounds become roughly constant when $m_\phi<q\approx \sqrt{2} \mu_N v_\odot$, where $q$ is the typical value of the transferred momentum in a nuclear recoil, with $\mu_N$ the reduced DM-nucleus mass and $v_\odot$ the Sun's velocity in the galactic frame.

\item \textbf{Cosmology:} If the dark photon is abundant in the early universe and decays into SM particles, predictions of the Big Bang nucleosynthesis (BBN) may be adversely affected. To avoid this possibility we can demand that the mediator decays before BBN begins, i.e. the decay rate of the dark photon is $\Gamma_\phi \approx \alpha_\text{em} m_\phi \epsilon^2/3 > 1\, \text{s}^{-1}$.

For non-thermal DM the cosmic microwave background (CMB) constrains the dark matter coupling $\alpha_X <0.17 (m_X/\text{TeV})^{1.61}$ from the imprints annihilation products would leave in the CMB. This upper limit is a fit of the results in \cite{Slatyer:2015jla} obtained by \cite{Feng:2016ijc}. The couplings we consider are always smaller than this upper limit.

\end{enumerate}

\begin{figure}[h!]
\includegraphics[width=.33\textwidth]{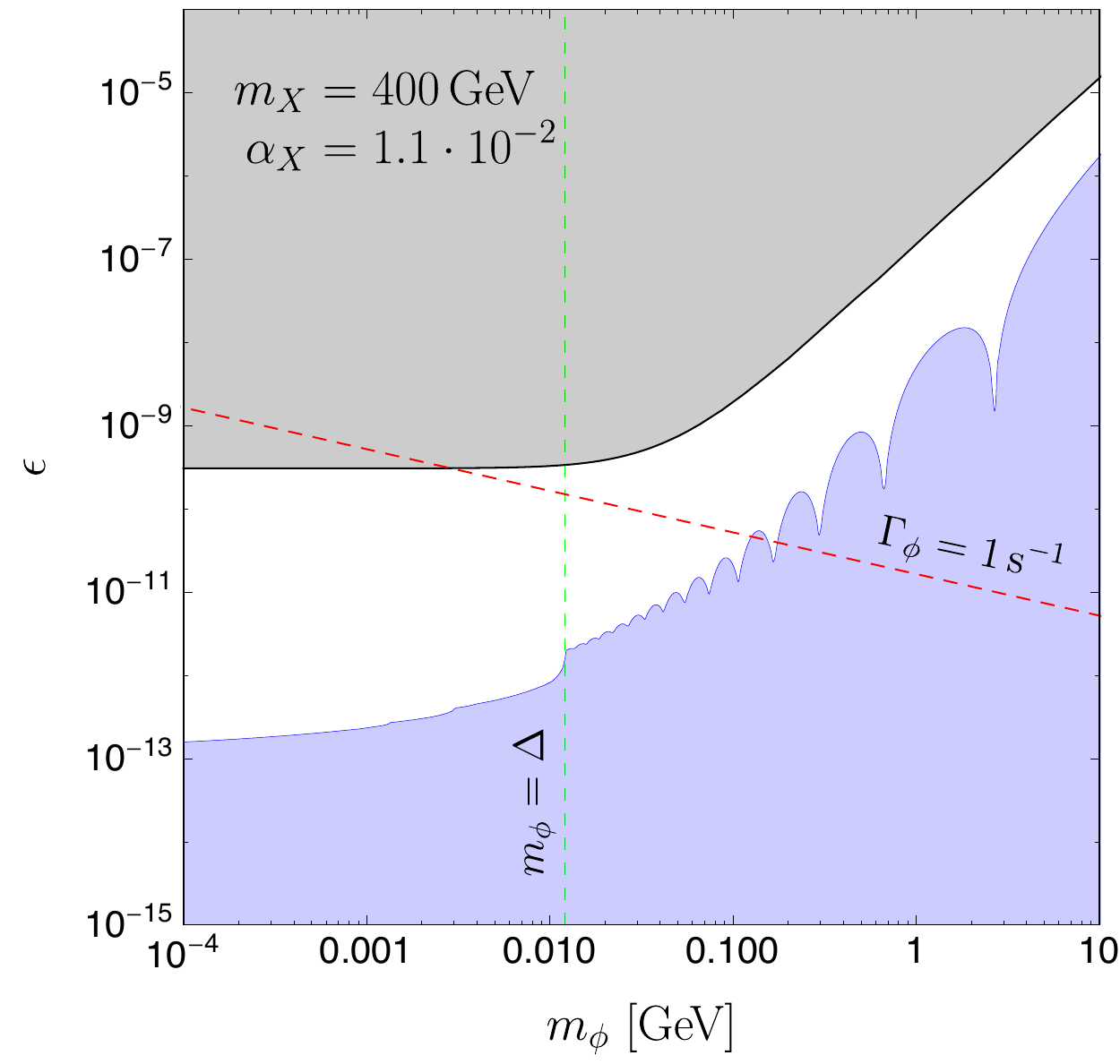}
\includegraphics[width=.33\textwidth]{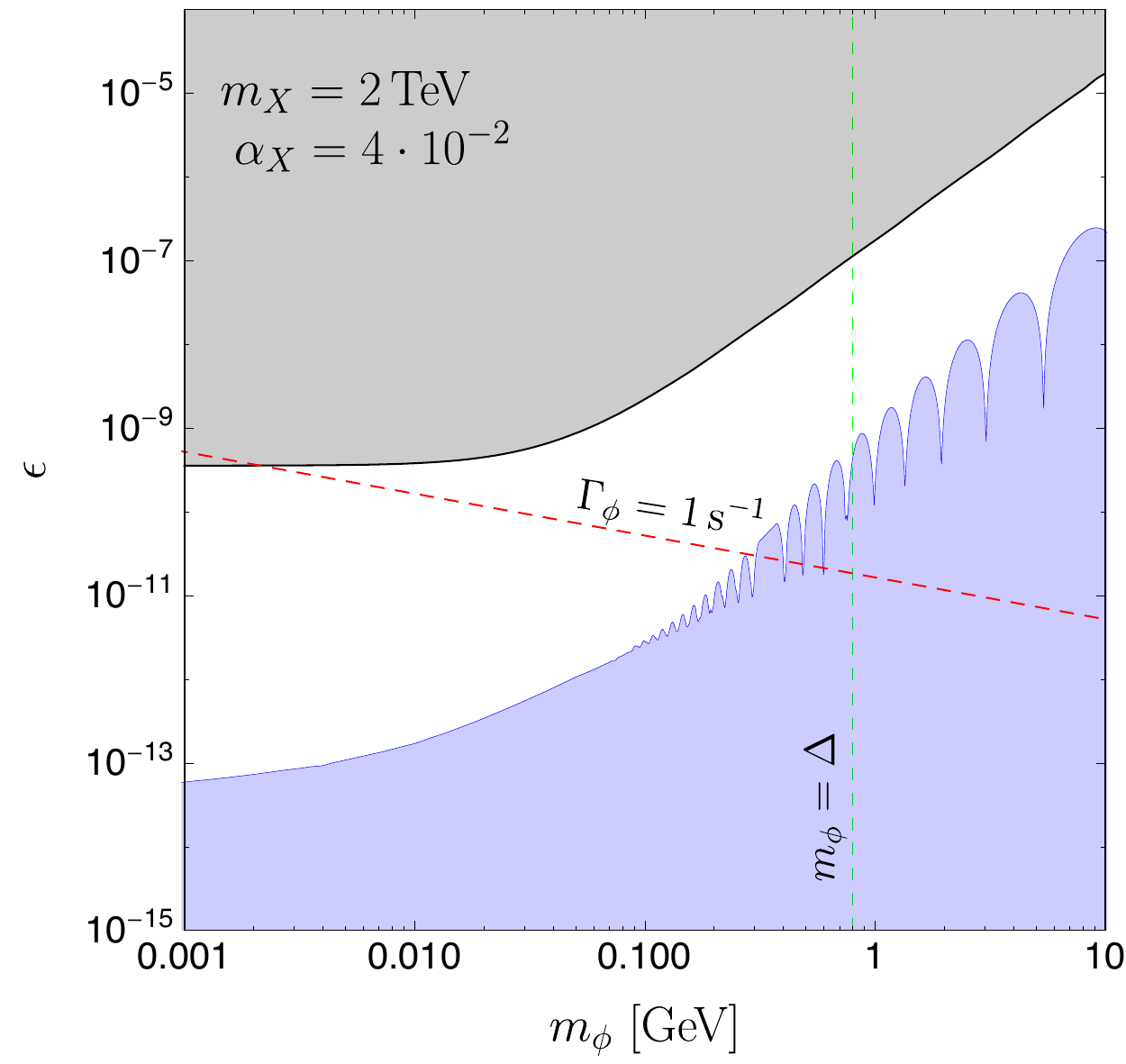}
\includegraphics[width=.33\textwidth]{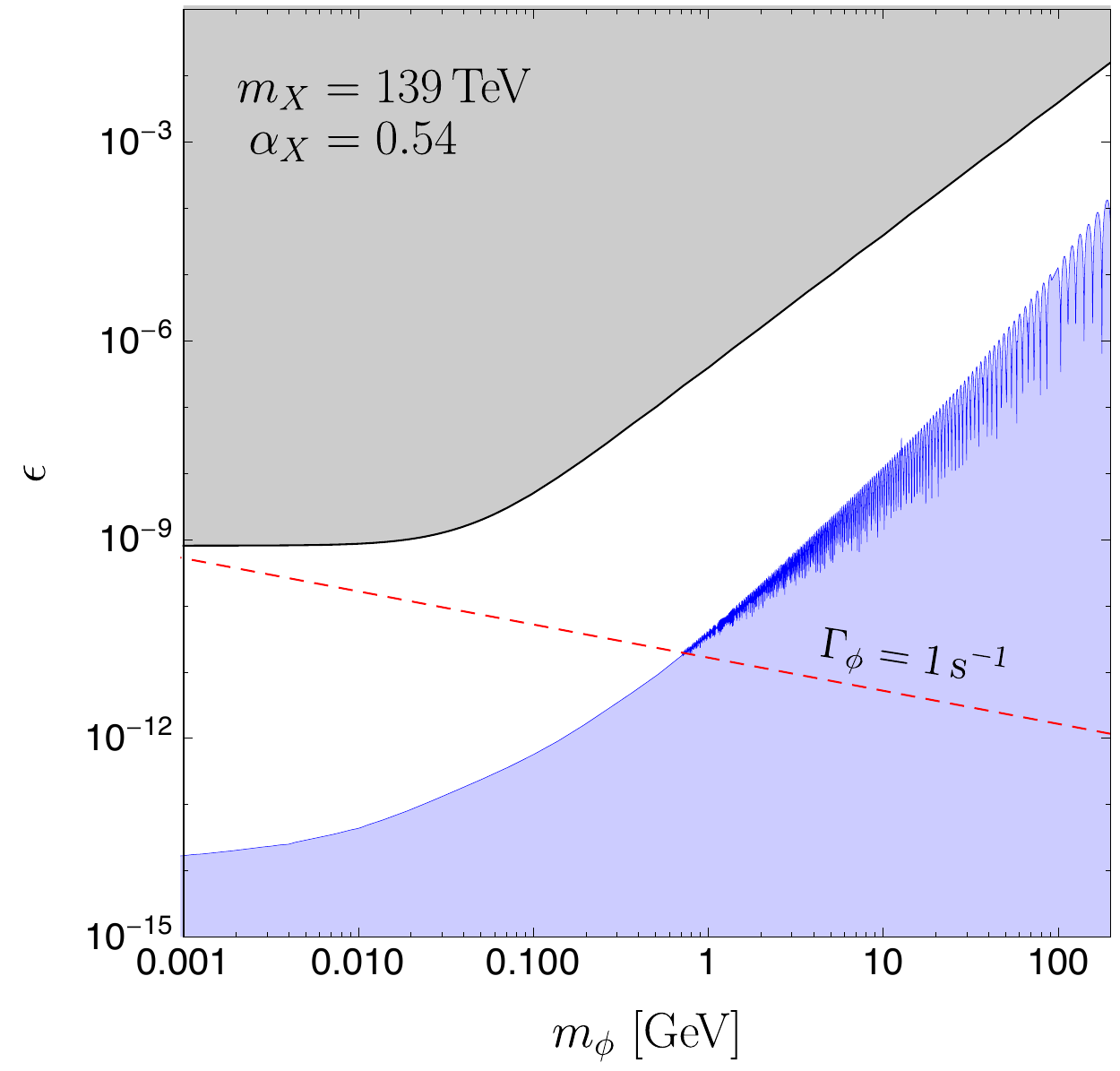}
\caption{Constraints on $\epsilon$ as a function of $m_\phi$. The gray contours are excluded by the LUX 2013 data at 90\% confidence level. The blue contours signify the region where the DM population in the Sun has not reached its steady state yet. Below the red line, the mediator decay time is longer than one second and could potentially be at odds with big bang nucleosynthesis. The vertical green lines show the binding energy of the ground state, when $m_\phi>\Delta$ the recombination rate is suppressed by $\epsilon^2$.}
\label{Fig:DD+SS constraints}
\end{figure}
To illustrate the behaviour of the signal, we choose three dark matter masses as benchmark points.  For each benchmark the mediator mass is varied in the region compatible with efficient bound state formation in the Sun. The benchmark points are chosen to be:

\begin{enumerate}[label=(B\arabic*)]
\item For our first benchmark we choose the (nearly) lightest DM mass where recombination can take place with DM self-interaction within the range solving the collisionless DM problems (see Fig.~\ref{Fig: Zurek Figure}): $m_X=400$ GeV and $\alpha_X =1.1 \cdot 10^{-2}$. The  mediator mass is chosen to be $m_\phi = 10$ MeV, slightly below the binding energy of the ground state which is $\Delta \approx 12$ MeV. The smallest kinetic mixing parameter for which the mediator decay is faster than one second is $\epsilon_\text{min} \approx 1.6 \cdot 10^{-10}$, while the maximum allowed by direct detection is $\epsilon_\text{max} \approx 3.3 \cdot 10^{-10}$.

\item Our second benchmark point corresponds to the point where the energy threshold of AMS-02 ($\sim 0.5$ GeV) is close to the maximum positron energy from a decaying recombination photon. This corresponds to $m_X=2$ TeV and $\alpha_X=4 \cdot 10^{-2}$, for which the ground state binding energy is $\Delta = 0.8$ GeV. We choose three values of the mediator mass: (a) $m_\phi = 5$ MeV, (b) $m_\phi = 15$ MeV and (c) $m_\phi = 70$ MeV.

\item For the last benchmark point we choose the heaviest DM particle allowed by unitarity of scattering. We adopt the value reported in \cite{vonHarling:2014kha} which is $m_X=139$ TeV and $\alpha_X=0.54$. Again we choose three values of the mediator mass: (a) $m_\phi = 2$ MeV, (b) $m_\phi = 20$ MeV and (c) $m_\phi = 100$ MeV. The value of $n_\text{max}$ is very large for this benchmark. However, Eq.~\ref{Eq: Recombination rate when it's large} is well satisfied within the range of the chosen parameters of this benchmark and  therefore  we use it for the recombination cross section in this case. 
\end{enumerate}
 In table \ref{Tab: Benchmark summary table} of appendix \ref{Sec: Benchmark summary appendix}   we summarize  key quantities  for each benchmark.


\section{Kinematics of mediator decay} \label{Sec: Kinematics of mediator decay}
In this section we review the formalism necessary to extract the positron spectrum from DM annihilating in the Sun. It is well known that the received gamma-ray flux emerging from the decay of a monochromatic (single energy) particle to two photons has  
 a characteristic box-like  spectrum  (see e.g. \cite{Ibarra:2012dw}), i.e the  gamma-ray flux is independent of energy. Alternatively, if the mediator is a Dirac fermion with chiral interactions, the spectrum form a triangle/trapezoid \cite{Ibarra:2016fco}. In our scenario dark photons are produced through either direct annihilation of DM particles or via recombination and subsequent darkonium decay. An addittional population is produced because a dark photon is emitted each time a recombination takes place inside the Sun. Dark photons created through direct annihilations, $p$-darkonium state decays and recombination are monochromatic, whereas dark photons originating from $o$-darkonium state decays are distributed according to Eq.~\ref{Eq: Ore-Powell spectrum}. Decays of monochromatic dark photons to positron-electron pairs produce a box-like spectrum  similar to the one of the gamma-rays. 

In principle, the energy of recombination photons depends on the excitation of the formed bound state. As already discussed, we make the simplifying approximation that all recombination photons have the ground state energy. As a further simplification, we neglect all dark photons, which may be emitted in transitions between darkonium states. For light mediators ($<2m_\mu$) the branching into electrons is almost 100\%. The  signal is for the most part heavily boosted, and therefore clearly directed towards the Sun. For highly energetic positrons, the bending in the magnetic field of the Sun will be small. Since the lightest DM particle we consider has a mass of $400$ GeV, the typical energy of positrons will be quite large. The positrons resulting from decays of recombination photons will in general be less energetic and thus experience a stronger effect. For simplicity we neglect all effects from the solar magnetic field.

\begin{figure}[h!]
\centering
\includegraphics[width=.6\textwidth]{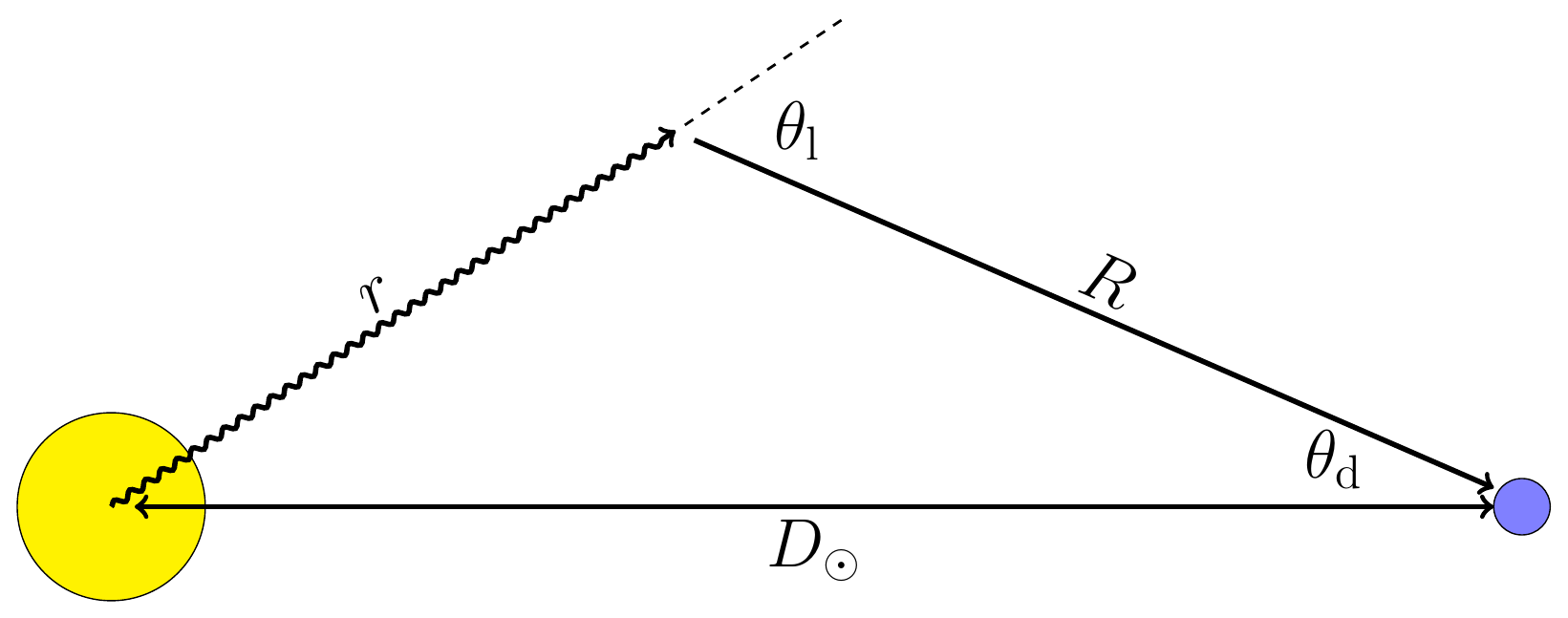}
\caption{Geometry of mediator decay. Dark photons are emitted from the Sun, and decay to positron electron pairs after distance $r$. Positrons reach the Earth with an angle $\theta_\text{d}$ with respect to the center of the Sun.}
\label{Fig:geometry}
\end{figure}

We now want to extract the flux of positrons in a solid angle directed towards the Sun. When the mediator is more boosted than the positrons there exists a maximum detector angle $\theta_\text{d}^\text{max}$, beyond which no positrons reach the Earth; see figure \ref{Fig:geometry} for definitions of lengths and angles. The flux  is described by the integral
 \begin{equation}
 	\phi_+ = -\int_0^{\theta_\text{d}^\text{max}} d \theta_\text{d}\sin\theta_\text{d} \frac{dN_+}{dAdtd\cos \theta_\text{d} dE_\text{d}},\label{Eq: flux}
 \end{equation}
where $\theta_\text{d}^\text{max} = \theta_\text{l}(E_\text{d})$ is the maximal detector angle for a specific detector energy $E_\text{d}$ (it is easy to understand why $\theta_\text{d}^\text{max} = \theta_\text{l}$ since two angles in a triangle sum to less than $\pi$, i.e. $\pi-\theta_\text{l} + \theta_\text{d}<\pi$). When comparing the flux with AMS-02 we smear out the signal in the regions where the maximum angle is smaller than experimental resolution, i.e. $\theta_\text{d}^\text{max}<\theta_\text{AMS}(E_\text{d})$, where $\theta_\text{AMS}(E_\text{d}) = \sqrt{(5.8^\circ)^2/(E_\text{d}/\text{GeV})+(0.23^\circ)^2}$ is the angular resolution of AMS-02~\cite{Gallucci:2015fma}.
In order to determine the integrand in Eq.~\ref{Eq: flux} we estimate the flux of positrons with energy $E_\text{d}$ arriving at a detector close to the Earth integrating over the  volume $2 \pi R^2 d\cos \theta_\text{d} dR$
\begin{equation}
\label{N1}
\frac{dN_+}{dAdt dE_\text{d}} = \int_0^\infty dR \int_0^\infty dE_\phi \frac{dN_\phi}{dVdtdE_\phi} \frac{1}{2\pi R^2}\frac{d\Gamma}{ d\cos \theta_\text{l}} (2 \pi R^2 d\cos \theta_\text{d}) \delta(E_\text{d}-E_+) \Theta(r-R_\odot),
\end{equation}
where $E_+$ is the positron energy, $d\Gamma/d\cos\theta_\text{l}$ is the fraction of positrons emitted at a particular  angle, such that it reaches the detector with energy $E_+$. Rearranging Eq.~\ref{N1} gives the integrand of Eq.~\ref{Eq: flux} \footnote{We note that our formula disagrees with Eq.~21 of \cite{Ajello:2011dq}.}
\begin{equation}
\frac{dN_+}{dAdtd\cos \theta_\text{d} dE_\text{d}} = \int_0^\infty dR \int_0^\infty dE_\phi \frac{dN_\phi}{dVdtdE_\phi} \frac{d\Gamma}{d\cos \theta_\text{l}} \delta(E_\text{d}-E_+) \Theta(r-R_\odot).
\label{Eq:Positron flux}
\end{equation}
 The last term is a Heaviside function that excludes positrons created inside the Sun. The first term in the integrand is the number of mediator decays per time, volume and energy $E_{\phi}$ at  a distance $r$ from the center of the Sun. It can be written as
\begin{equation}
\frac{dN_\phi}{dVdtdE_\phi} = \frac{e^{-r/L}}{L} \frac{\Gamma_\phi(E_\phi)}{4\pi r^2},
\label{Eq:mediator flux}
\end{equation}
where $L$ is the decay length of $\phi$ in Eq.~\ref{Eq: decay length} and $\Gamma_\phi(E_\phi)$ is the rate of emitted mediators per energy $E_\phi$. The form of $\Gamma_\phi$ changes depending on the specific DM model. If only direct annihilation is taken into account the rate is just $\Gamma_\phi(E_\phi) = C_\text{cap} 2\delta(E_\phi -m_X)$. In our case $\Gamma_\phi$ is significantly more involved 
\begin{equation}
	\Gamma_\phi(E_\phi) = C_\text{cap} \left( k_o\frac{dN_\phi^{(o)}}{dE_\phi}  + k_p\frac{dN_\phi^{(p)}}{dE_\phi} + k_\text{ann}\frac{dN_\phi^{(\text{ann})}}{dE_\phi} + (k_o+k_p)\frac{dN_\phi^{(\text{rec})}}{dE_\phi}\right),
	\label{Eq:energy rate}
\end{equation}
where $k_i$ is the fraction of DM particles that are converted to photons through either direct annihilation or decay of the $o$- and $p$-darkonium states. By inspection of the Boltzmann equations Eqs.~\ref{Eq:Boltzmann eqs a}-\ref{Eq:Boltzmann eqs c} and using the steady state values of Eq.~\ref{Eq: Steady states}, we get $k_o = 3 k_p = 3C_\text{rec}/(4(C_\text{rec}+C_\text{ann}))$ and $k_\text{ann} = C_\text{ann}/(C_\text{rec}+C_\text{ann})$. $dN_{\phi}^{(i)}/dE_{\phi}$ are given by Eqs.~\ref{Nann}, \ref{Ndp},  \ref{Eq: Ore-Powell spectrum} and \ref{Nrec}.
 The factor $1/4\pi r^2$ in Eq.~\ref{Eq:mediator flux} assumes that the mediators are radiated isotropically  from the Sun. 
 Eq.~\ref{Eq:Positron flux} can be simplified significantly by noting that $\phi$ decays isotropically in its center of mass (cm) frame, and that the energy of a positron in the lab frame is only a function of the decay angle in the cm frame $\theta_\text{cm}$
\begin{equation}
E_+ = \frac{\gamma_\phi m_\phi}{2} (1+v_+v_\phi \cos \theta_\text{cm}).
\label{Eq: Eplus}
\end{equation}
Here $v_\phi = (1-m_\phi^2/m_X^2)^{1/2}$, $v_+ = (1-4m_e^2/m_\phi^2)^{1/2}$ and $\gamma_\phi = (1-v_\phi^2)^{-1/2}$. Each angle $\theta_\text{cm}$ corresponds to a particular angle in the lab frame
\begin{equation}
\cos \theta_\text{l} = \frac{\gamma_{\phi}\left(\cos\theta_\mathrm{cm}  + \alpha \right)}{\sqrt{\gamma_{\phi}^2\left(\cos\theta_\mathrm{cm} + \alpha \right)^2+\sin^2\theta_\mathrm{cm}}}\, ,
\label{Eq: specrel}
\end{equation}
where $\alpha =v_\phi / v_+$. Combining Eq.~\ref{Eq: Eplus} and \ref{Eq: specrel} we can write $\theta_l(E_+)$. When $\alpha \geq 1$, there is a maximum angle $\theta_\text{l}^\text{max} \leq \pi/2$, such that for each $\theta_\text{l}<\theta_\text{l}^\text{max}$ there are two corresponding angles in the center of mass frame and thus two positron energies. We note, that the width of the positron energy spectrum in the lab frame is given by $\Delta E_{+} = \gamma_\phi m_{\phi} v_+ v_\phi$. We can now rewrite the last terms of the integrand by applying the chain rule and the identity $\delta[f(x)] = \delta(x-x_0)/|f'(x_0)|$
\begin{align}
 \frac{d\Gamma}{d\cos \theta_\text{l}} \delta(E_\text{d}-E_+)&=  \frac{d\Gamma}{d\cos \theta_\text{cm}} \left| \frac{d\cos \theta_\text{cm}}{d\cos \theta_\text{l}} \right| \delta(E_\text{d}-E_+) \notag \\
  &=  \frac{d\Gamma}{d\cos \theta_\text{cm}} \left| \frac{dE_+}{d\cos \theta_\text{cm}}  \frac{d\cos \theta_\text{l}}{dR}\right|^{-1} \delta(R-R_0) \notag \\
 &= - \frac{1}{\gamma_\phi m_{\phi} v_+ v_\phi} \frac{r^3}{\sin^2\theta_\text{d} D_\odot^2 } \delta(R-R_0),
 \label{Eq:angular part of flux}
\end{align}
where the absolute value in the first line ensures that a positive number of particles are emitted in the interval $d\cos \theta_\text{l}$, $D_\odot = 1$ a.u. $ = 1.50\cdot 10^{11}$ m is the distance to the Sun, $d\Gamma/d\cos\theta_\text{cm} = -1/2$ from isotropy and $dE_+/d\cos \theta_\text{cm} = \gamma_\phi m_\phi v_+ v_\phi /2$. In the last line we use $d\cos \theta_\text{l}/ dR = - \sin^2\theta_\text{d} D_\odot^2/r^3$ which follows from the geometric identity 
\begin{equation}
	\cos \theta_\text{l} = \frac{D_\odot \cos\theta_\text{d} - R}{r},
	\label{Eq: Cosine relations}
\end{equation}
with $r^2 = R^2 +D_\odot^2-2R D_\odot \cos\theta_\text{d}$.
When evaluating the $R$-integral in Eq.~\ref{Eq:Positron flux} it is useful to know $R_0$ in Eq.~\ref{Eq:angular part of flux} which for a particular detector energy (or equivalently a particular $\theta_\text{l}$) can be found from Eq.~\ref{Eq: Cosine relations}
\begin{equation}
R_0 = \left(\cos\theta_\text{d}- \sin\theta_\text{d} \cot\theta_\text{l} \right) D_\odot.
\end{equation}
For a fixed detector angle, the distance $r$ is only a function of $R$. For evaluating the angular integral it is also useful to note the quantity $r_0 = r(R_0)$ which is
\begin{equation}
r_0 = \sin\theta_\text{d}\csc\theta_\text{l}  D_\odot.
\end{equation}
After trivially integrating over $R$ (by using the delta function) Eq.~\ref{Eq:Positron flux} can be written as
\begin{equation}
\frac{dN_+}{dAdtd\cos \theta_\text{d} dE_\text{d}} = -\int_0^\infty dE_\phi\frac{\Gamma_{\phi }( E_{\phi})\csc\theta_\text{l}}{4\pi E_{\phi} v_+ v_\phi D_\odot L}\, \csc\theta_\text{d} e^{-\sin\theta_\text{d}\csc\theta_\text{l}  D_\odot/L} \Theta(\sin\theta_\text{d}\csc\theta_\text{l}  D_\odot-R_\odot) ,
\label{Eq:Positron flux Rintegrated}
\end{equation}
where $E_{\phi}=\gamma_\phi m_{\phi}$ is the energy of the mediator and $\Gamma_{\phi}( E_{\phi})$ is the rate Eq.~\ref{Eq:energy rate} . In the case where $\alpha > 1$, $\theta_\text{d}^\text{max} = \theta_\text{l}(E_\text{d}) < \theta_\text{AMS}(E_\text{d}) \ll 1$, $\phi_+$ becomes approximately
\begin{equation}
\phi_+= \int_0^\infty dE_\phi \frac{\Gamma_{\phi}(E_{\phi})}{4\pi E_{\phi} v_+ v_\phi D_\odot^2}\,  \left(e^{-R_{\odot}/L} -e^{-D_{\odot}/L}\right)\, ,
\label{Eq:Positron flux integrated}
\end{equation}
where the last term is  the fraction of dark photons with energy $E_{\phi}$, which decays between the surface of the Sun and the Earth. The three processes; annihilation, recombination and para-decay contribute to $\Gamma_{\phi}( E_{\phi})$ in the form of delta-functions. In these cases, when Eq.~\ref{Eq:Positron flux integrated} is valid, the signal is independent on $E_d$. Hence, when the mediator is more boosted than the electron-positron pair it decays into, the corresponding spectrum is approximately a box. However, for $\alpha < 1$, $\cos\theta_\text{l}$ takes values in its full range, and for energies $E_\text{d}$ where $\theta_\text{l}(E_\text{d}) > \theta_\text{AMS}(E_\text{d})$, the signal is spread over more than one angular bin of AMS. In this case, we should also subtract positrons emitted  behind the Sun. The width of the spectrum is given by $\Delta E_{+} = \gamma_\phi m_{\phi} v_+ v_\phi$. The height of the signal scales linearly with the capture rate, and inversely with the width of the spectrum. Furthermore, Eq.~\ref{Eq:Positron flux integrated} shows the dependence on the decay length $L$. For $L\ll R_{\odot}$ all the decays happen within the Sun, whereas for $L\gg D_{\odot}$ all decays occur past the Earth. From  Eq.~\ref{Eq: decay length} and \ref{Eq: capture rate}, we get that $L\propto \epsilon^{-2}$, $C_\text{cap}\propto \epsilon^2$, and conclude that for each pair $\gamma_\phi, m_{\phi}$ there is an optimum value of $\epsilon$ that maximizes the detected positron flux.
When combining the positron spectrum from mediators emitted in different processes, the signal will not scale uniformly with $\epsilon$.  
\begin{figure}[h!]
	\centering
	\includegraphics[width=.5\textwidth]{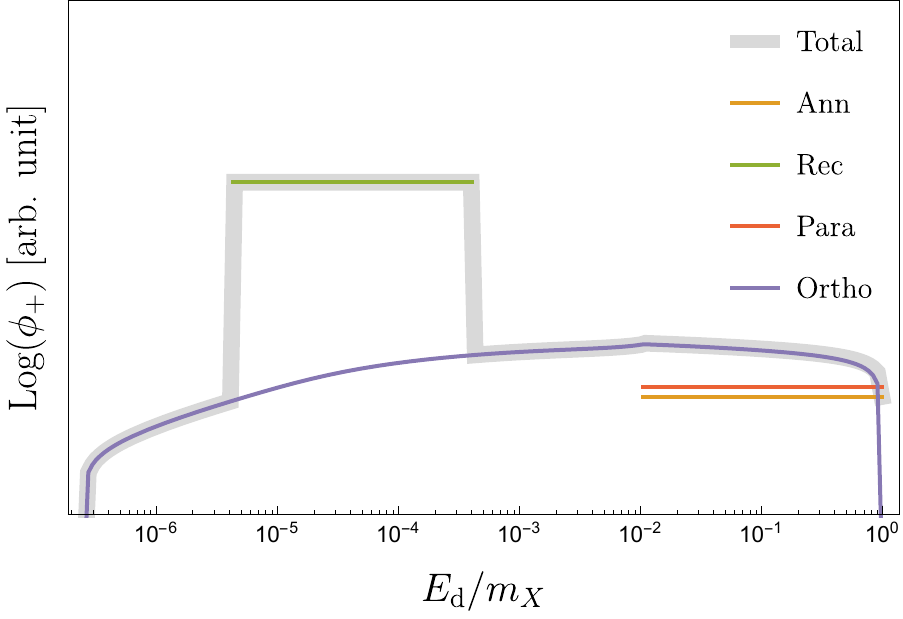}
	\caption{Sketch of the combined positron spectrum. The thick, gray curve depicts the total signal, whereas the colored curves are the contributions from annihilation, recombination and decays of $p$- and $o$-bound states.}
	\label{Fig:SplitSignal}
\end{figure}
Since $\Gamma_\phi(E_\text{d})$ has four contributions, where three are proportional to delta-functions and the last is proportional to the Ore-Powell spectrum, the final flux will be a sum of three boxes of unequal height and width, along with a fourth contribution which depends on $E_\text{d}$. From the width of each box, we get information about the energy carried by the dark photon. Particularly, the dark matter mass, $m_X$, and the binding energy, $\Delta $. The height and shape of the signal will, in turn, determine the kinetic mixing, $\epsilon$, and the mediator mass, $m_\phi$. In Fig.~\ref{Fig:SplitSignal}, we show a sketch of the combined positron spectrum with color-coded composition.

\section{Results} \label{Sec: Results}
\begin{figure}[h!]
\centering
    \includegraphics[width=.45\textwidth]{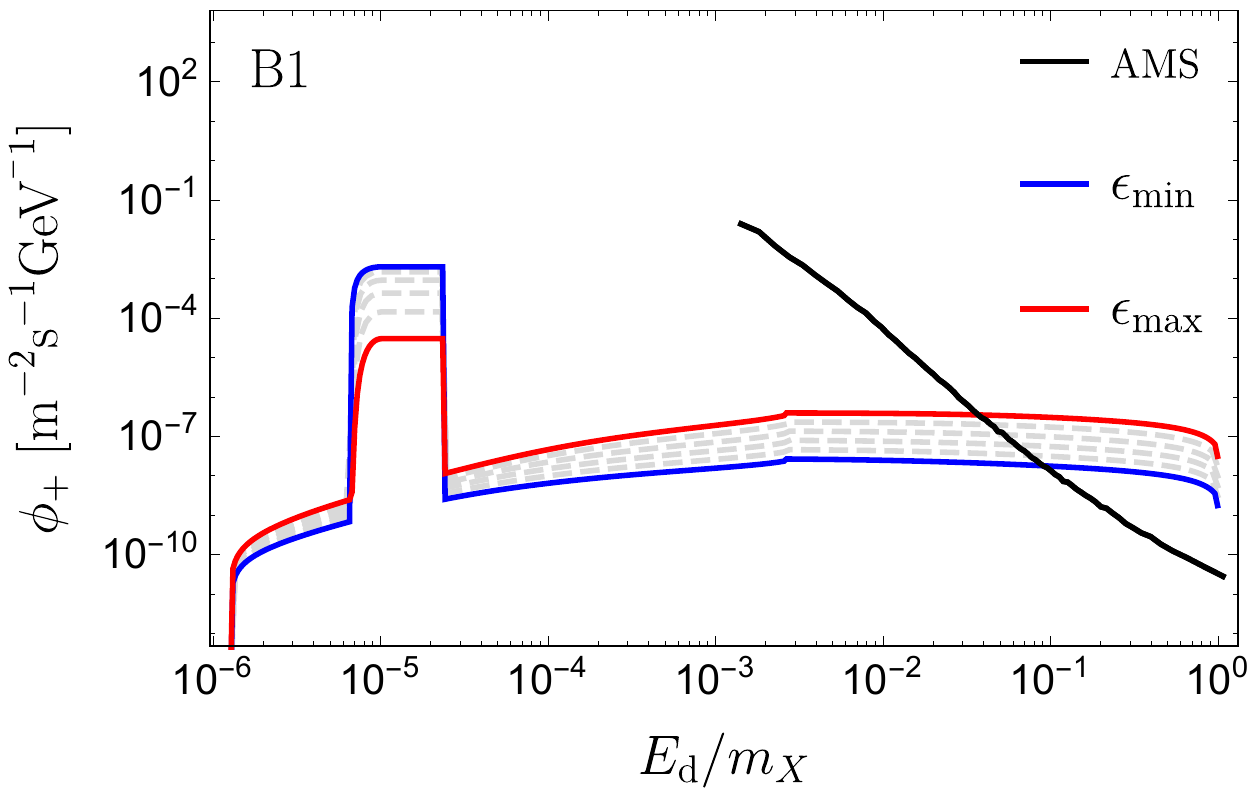}
    \includegraphics[width=.45\textwidth]{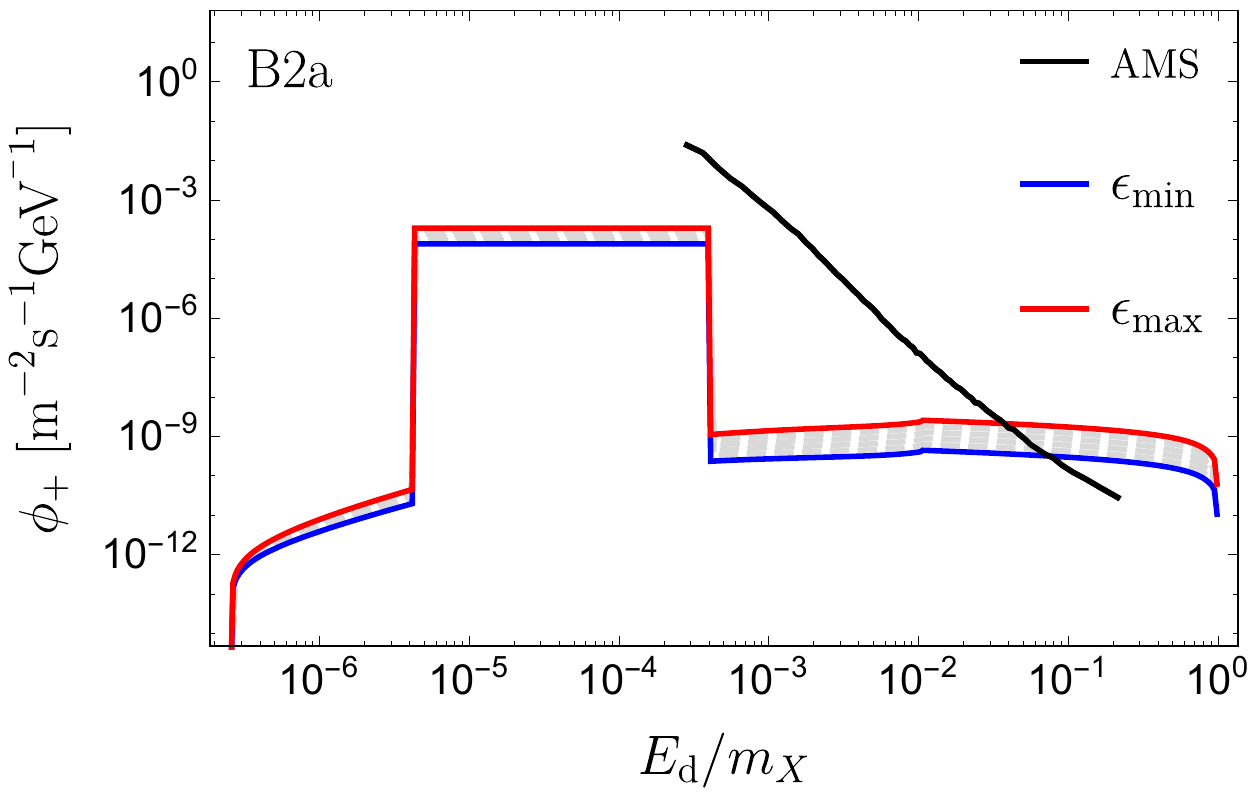}
	\includegraphics[width=.45\textwidth]{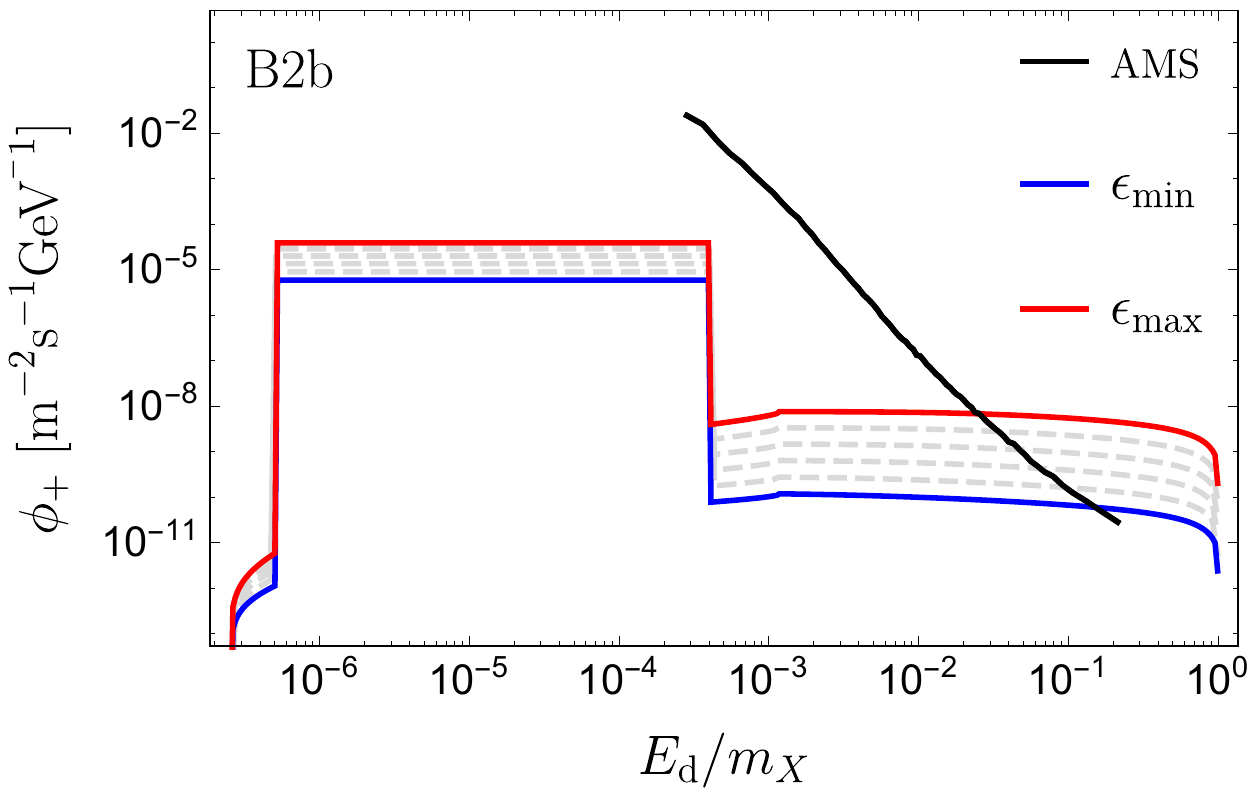}
	\includegraphics[width=.45\textwidth]{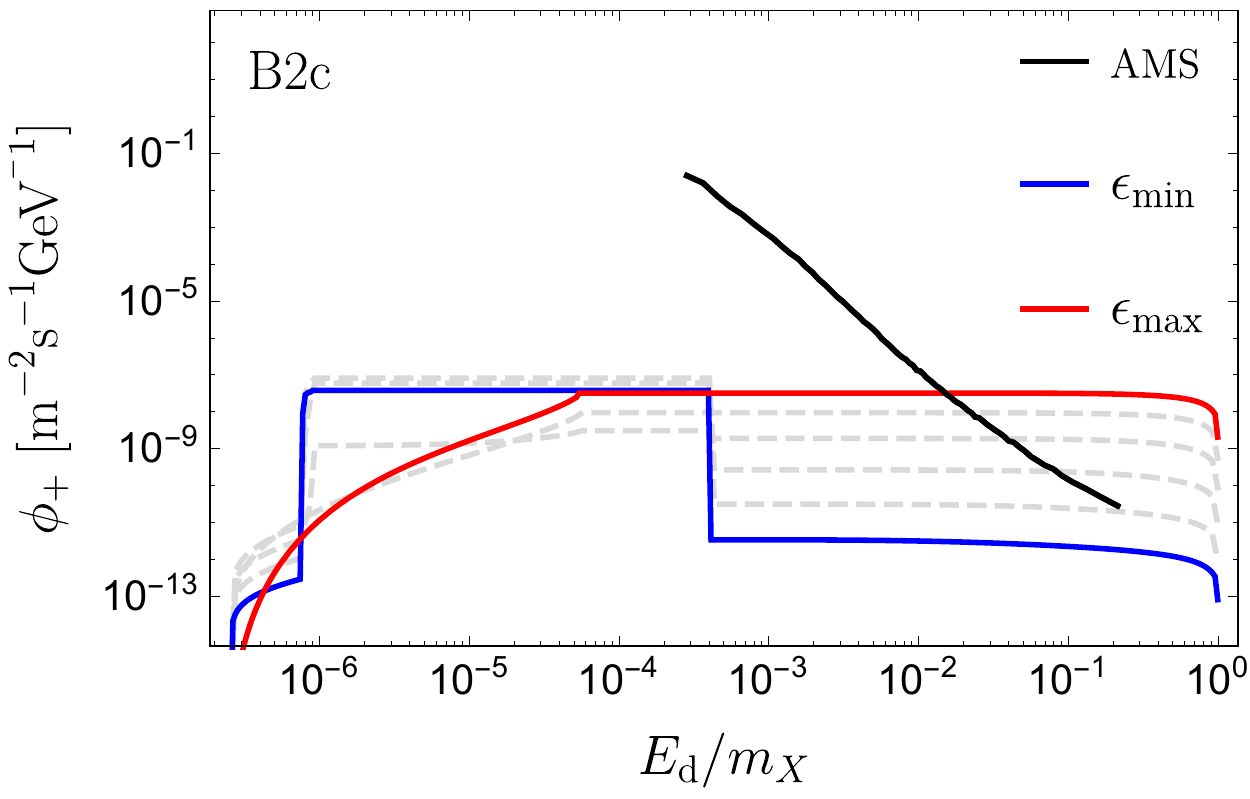}
	\includegraphics[width=.45\textwidth]{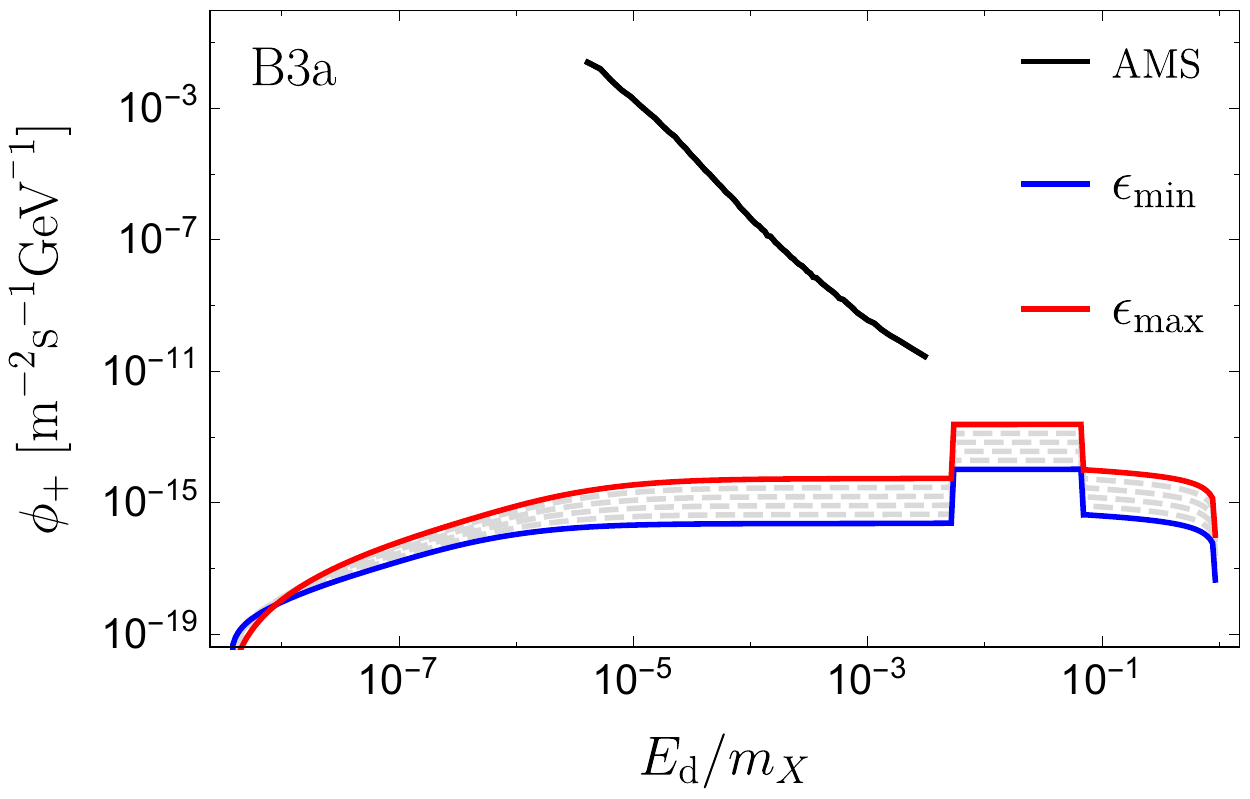}
	\includegraphics[width=.45\textwidth]{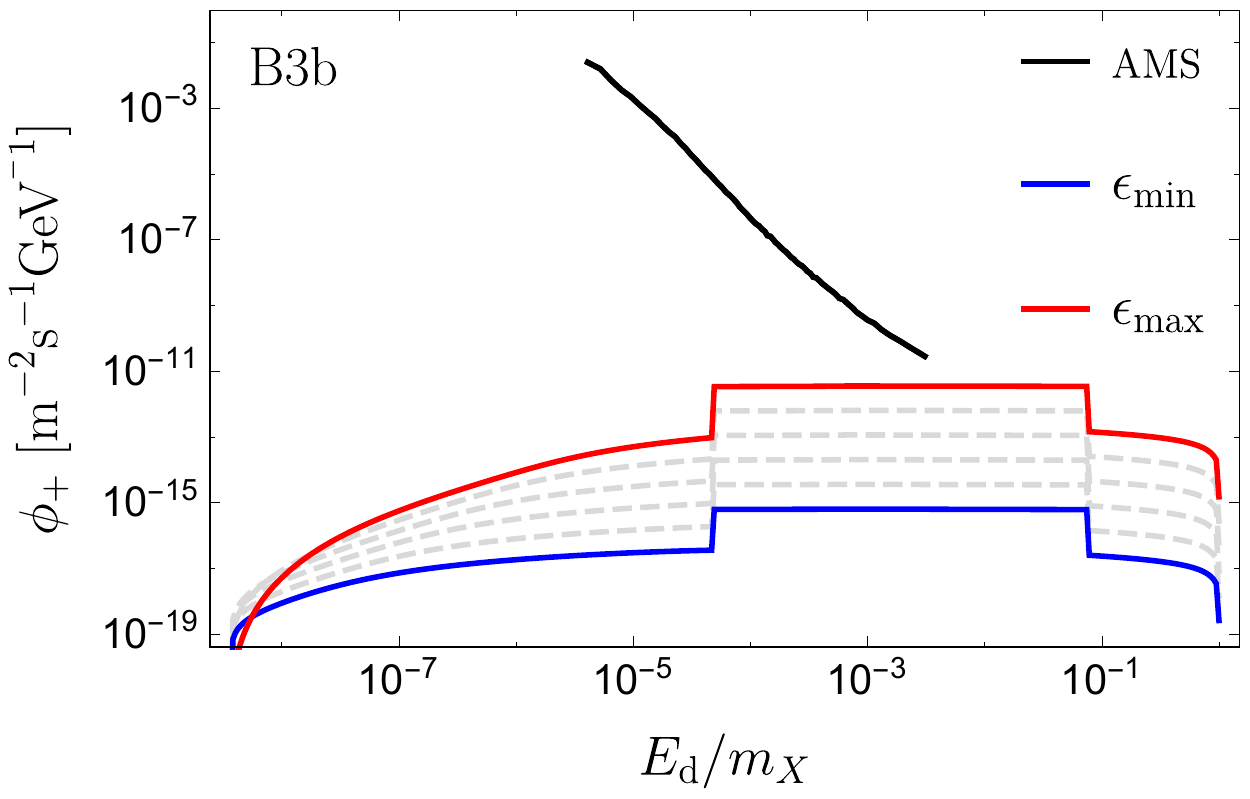}
	\includegraphics[width=.45\textwidth]{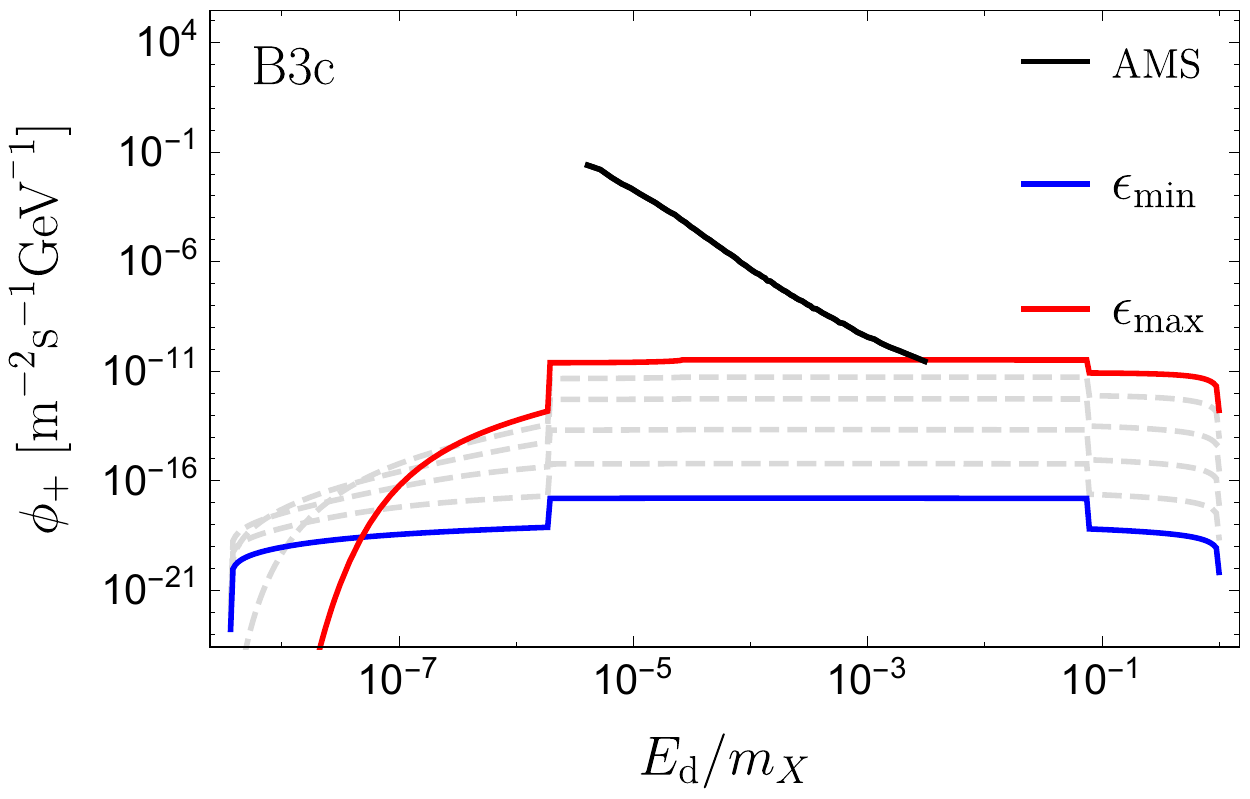}
	\caption{These figures show the solar positron spectra from DM annihilation. The red (blue) curve assumes the maximum (minimum) allowed $\epsilon$ for each benchmark. The solid black curve is the uncertainty in the isotropic positron flux measured by AMS-02. }
	\label{Fig: LogLog Spectra}
\end{figure}

In this section we present the spectra of positrons from DM annihilating in the Sun using the formalism of the previous section. The main results of the paper are contained in Fig.~\ref{Fig: LogLog Spectra}. This figure shows the positron flux from the Sun as a function of energy for our benchmark parameters. In Fig.~\ref{Fig: LogLog Spectra} we have superimposed the $1\sigma$ uncertainty contours of the isotropic positron flux measured by AMS-02. Since AMS-02 measures an isotropic positron flux and the DM signal is anisotropic and directed towards the Sun, the signal must be smaller than the uncertainty on the isotropic flux measurement. We indeed find that this uncertainty is exceeded by the DM signal from the Sun in some cases. 
 We will however refrain from placing new constraints on the parameter space, since we lack data on the positron flux in the angular bins directed towards the Sun. Specifically, to place new constraints on the model's parameter space we need the measured solar positron flux $\phi_\odot$ and the associated statistical error $\delta\phi_\odot$. We would then require the anisotropic DM signal $\phi_\text{DM}$ to be lower than the difference in the solar and isotropic fluxes, i.e. $\phi_\text{DM} < \phi_\odot-\phi_\text{iso} + \delta\phi_\text{iso} + \delta\phi_\odot$, where $\phi_\text{iso}$ and $\delta\phi_\text{iso}$ are the isotropic positron flux and the associated error. To be conservative, we add the statistical errors on the isotropic and solar fluxes.

In Fig.~\ref{Fig: LogLog Spectra}  we see that for all benchmarks the spectra are dominated by the signal from the recombination photons and the Ore-Powell (ortho-decay) photons. The dependence of the spectrum on $\epsilon$ is subtle. On the one hand
large values of $\epsilon$ increase the capture rate and potentially the positron signal. On the other hand large $\epsilon$ values  decrease the decay length. For high energetic dark photons e.g. from direct annihilation, the decay length is sufficiently large so the overall effect of increasing $\epsilon$ corresponds to an increase in the overall positron spectrum. On the contrary for not so energetic dark photons like those of recombination, the decay length might not be so large to make it out of the Sun, thus increasing the value of $\epsilon$ reduces the positron flux.  This is depicted e.g. in B1. As we shall see this is not a universal feature i.e.  other benchmark points have larger positron fluxes for larger values of $\epsilon$ even in the recombination part of the spectrum.

In panel B1, the recombination signal lies at much lower energies than those probed by AMS-02. The recombination photons have relatively small boost factors, hence the signal is not entirely box-shaped. The right side of the recombination box signal is from positrons emitted forward in the lab frame with a small angle $\theta_\text{l}(E_\text{d}) < \theta_\text{AMS}(E_\text{d})$, such that AMS would detect the full signal within one angular bin. When going to the left side of the box-like signal, the angle $\theta_\text{l}(E_\text{d})$ increases and part of the signal exceeds $\theta_\text{AMS}(E_\text{d})$ leading to a lower detected flux, thus smearing the left side of the box.  For $\theta_\text{l}(E_\text{d})>\pi/2$, the positrons are emitted at distances $R > D_\odot$ and a significant part of the signal is obstructed by the Sun. The low boost factor of the recombination photon makes the signal decrease with increasing $\epsilon$ because the decays occur predominantly inside the Sun. Contrary to this, the signal from the highly boosted Ore-Powell photons increase with increasing $\epsilon$. The gray dashed curves indicate the flux for $\epsilon$ between $\epsilon_{\text{min}}$ and $\epsilon_{\text{max}}$. 
We note that the full range of allowed values for $\epsilon$ gives rise to a spectrum which exceeds the error on the isotropic signal from AMS-02 in the high energy end of the data. For B2a, $m_\phi \ll \Delta$ such that the recombination photons are highly boosted, and the whole spectrum is elevated when $\epsilon$ is increased. Again in this case, the high energetic part of the spectrum is above the AMS error. The recombination signal is now within the energy range of AMS, but more than an order of magnitude below the error. By increasing the mediator mass, the width of the recombination signal broadens, since $v_+ \to 1$, and $\epsilon_{\text{max}}$ is increased. For B2b, the high energetic Ore-Powell photons are still above the AMS error. When increasing the mediator mass further, we reach a point where the recombination signal resembles the one in B1, and similarly decreases with $\epsilon$. For B2c, the recombination signal is only visible in the lower end of the allowed range of $\epsilon$. Likewise, the lower end of the Ore-Powell spectrum is no longer increasing with $\epsilon$. In spite of this, the upper part of the spectrum rises above the AMS error for most of the allowed range for $\epsilon$.  When going to the largest dark matter mass, shown in panel B3a-c, we get a similar series of pictures to B2a-c. However, now the binding energy is clearly within the positron energy range for AMS. Furthermore, the effect on the width of the recombination signal is more pronounced when comparing B3a with B3c. For B3a-b, the positron signal is below the error on AMS, while for B3c the positron isotropic uncertainty of AMS-02 barely touches the positron signal from dark recombination photons (for the maximum possible value of $\epsilon$). This is quite interesting  because in this case AMS-02 will be able to probe the recombination dark photons before the annihilation ones.    

Apart from positron spectra, there are identical electron spectra. Furthermore, the dark photon can decay into three real photons, although with a suppressed branching  $\sim \text{Br}(\phi \to 3\gamma)\simeq 0.02 \alpha^3  = 8\cdot 10^{-9}$~\cite{Glover:1993nv} in the case $m_e\ll m_\phi$. 

Our results show that AMS-02 can potentially discover a whole spectrum of positrons produced from dark photon decays outside the Sun. Alternatively this can be used to constrain DM models with dark photons. AMS-02 is only directed towards the Sun 1/80th of its livetime~\cite{Feng:2016ijc}. The uncertainty in the subset of the data from the Sun is likely much larger than the isotropic error. Collection of more data and knowledge of the measurements in the bins directed towards the Sun would potentially impose strong constraints.

\section{Conclusions} \label{Sec: Conclusions}
In this paper we studied the positron spectrum produced by decaying dark photons that originate in the Sun. We consider dark photons that are produced inside the Sun via direct annihilation of $X\bar{X}$, decay of bound states $X\bar{X}$ and bound state formation. These photons decay to positron-electron pairs and those decays that take place outside the Sun can potentially create a positron flux on Earth, that could be detected e.g. by AMS-02.
We demonstrated that the spectrum has distinct features as a function of the positron energy that could distinguish it from any other astrophysical background of positrons. These features are due to contributions from different types of dark photons i.e. coming from direct annihilation, decay of para- or ortho-darkonium and recombination. More importantly we find that there is parameter space where AMS-02 or equivalent experiments would be able to pick up the contribution from the dark recombination photons before they have a chance to observe the positrons that come from dark photons produced by the typical direct annihilation of DM inside the Sun. This result appears to be surprising since in principle recombination photons are fewer and less energetic than those from direct annihilation. However as we have argued, the positrons produced from the decay of recombination photons span a much narrower energy band than the rest, leading to  higher numbers in lower energy bins. Possible discovery of a positron spectrum with the morphology of our Fig.~\ref{Fig:SplitSignal} not only can associate beyond any doubt the positron signal to its DM origin, but it could establish the type and the mass of the mediator, thus understanding how DM self-interacts.
\section*{Acknowledgements}
The CP$^3$-Origins centre is partially funded by the Danish National Research Foundation, grant number DNRF90. We would also like to thank Jordan Smolinsky for valuable correspondence regarding the solar capture rate.

\appendix
\section{Verification of approximations} \label{App: Verification of approximations}
When solving the Boltzmann equations inside the Sun we have neglected a number of effects. Below we discuss the various effects we have neglected. The approximations are verified in table~\ref{Tab: Probabilities}.
\begin{enumerate}
\item {\bf Darkonium ionization is negligible with respect to decay:}\\
The ionization cross section is related to the recombination cross section by the Milne-relation
\begin{equation}
\sigma_\text{ion} = \frac{m_X^2 v^2}{8\omega^2} \sigma_\text{rec} \approx \frac{m_X^2v_\text{th}^2}{4\Delta^2} \sigma_\text{rec},
\end{equation}
where $\omega = \Delta+m_X v^2/4 \approx \Delta$ is the energy of the photon. For our benchmark parameters, Eq.~\ref{Eq: Recombination rate when it's large} describes $\sigma_\text{rec}$ within an order of magnitude, so we use this expression in the following estimate. If the mean free path for a dark photon ionizing darkonium is much larger than the thermal radius where DM is concentrated in the Sun, then ionization from recombination photons can be neglected. 
The ionization mean free path is $\lambda_\text{ion} \sim 1/(\sigma_\text{ion} n_D)$ where $n_D$ is the number density of darkonium. We use the steady state darkonium number densities to estimate the probability ($P_\text{ion} =1-e^{-r_\text{th}/\lambda_\text{ion}}\approx r_\text{th}/\lambda_\text{ion}$) to ionize a darkonium particle before leaving the Sun. The values are given in table \ref{Tab: Probabilities}.
Since recombination photons and darkonium are produced in the ratio 1:1, ionization will be negligible if recombination photons are very likely to leave the Sun. One might worry that the larger number of dark photons from annihilation and darkonium decay can mitigate the small likelihood of ionizing a darkonium state such that ionization becomes relevant. However, these dark photons have higher energy ($\simeq m_X$) and their likelihood for ionizing darkonia is much smaller. As a result ionization remains completely negligible.

\item {\bf Only ground state darkonium is appreciably populated in the Sun}.\\
When an excited state of darkonium makes a transition to the ground state by emission of a Lyman series photon, this photon can resonantly excite another ground state. This process can only be efficient if the mean free path of the Lyman-$\alpha$ dark photon is small compared to the thermal radius. The excitation cross section near resonance is \cite{Peebles:1994xt}
\begin{equation}
\sigma_{\text{L}\alpha} = \frac{3\lambda^2}{8\pi}\frac{\gamma^2}{(\omega-\omega_0)^2 + \gamma^2/4}
\end{equation}
where $\lambda = 2\pi/\omega$, $\omega_0  = \Delta-\Delta_2 = 3\alpha_X^2 m_X/16$ is the photon energy, $\gamma$ is the decay rate of the $2p$ state. Exactly on resonance the cross section becomes
\begin{equation}
	\sigma_{\text{L}\alpha}=\frac{512 \pi}{3\alpha_X^4m_X^2}.
\end{equation}
 As before, we estimate the excitation probabilities by replacing $\sigma_\text{ion}$ with $\sigma_{\text{L}\alpha}$ and summarise in table \ref{Tab: Probabilities}. The conclusion is that the effect of exciting ground state darkonia by absorbsion of Lyman-$\alpha$ dark photons is negligible.

\item \textbf{Dark photons do not scatter before leaving the Sun.}\\
We now verify that the dark photons  escape from the Sun with a negligible probability of scattering on DM or solar SM particles. The mean free path  is $\lambda_\text{C}^{(T)} =  1/(n_T \sigma_\text{C}^{(T)})$, where $n_T$ is the number density of targets and $\sigma_\text{C}$ is the Compton scattering cross section on target $T$. If we take the target to be DM particles, we must have $\lambda_\text{C}^{(X)}\gg r_\text{th}$, whereas for electrons, we must have $\lambda_\text{C}^{(e)} \gg R_\odot$. Among the SM particles we consider electrons because with a mean density of $n_e \sim 10^{30}/$m$^3$ constitute the most efficient target. The $\phi$-electron Compton scattering at the energies of interest i.e. $\sqrt{s} \gg m_e$ is
\begin{equation}
\sigma_\text{C}^{(e)}=	\frac{2\pi \epsilon \alpha_X \alpha}{s} \log \frac{s}{m_e^2}.
\end{equation}
The largest  value for the cross section is achieved for $\phi$ produced in recombination where $\sqrt{s} = \Delta$.

For $\phi$ scattering off DM particles, we use the cross section
\begin{equation}
\sigma_\text{C}^{(X)}=\frac{8\pi  \alpha_X^2}{3 m_X^2}.
\end{equation}
Again we summarise the scattering probabilities in table \ref{Tab: Probabilities}.

\item \textbf{DM self-capture is negligible when the DM population in the Sun is at its steady state value.}\\
For DM self-capture to be negligible, the self-capture rate must be small compared to that on nuclei. In the steady state this requirement can be written as $C_\text{self-cap} N_X^\text{SS}\ll C_\text{cap}$. Equivalently, we must have $\tau_X \ll 1/C_\text{self-cap}$. Reference \cite{Feng:2016ijc} has verified this is true if only direct annihilations are taken into account. Since \cite{Feng:2016ijc} neglected recombination and $\tau_X = 1/\sqrt{C_\text{cap}(C_\text{ann}+C_\text{rec})}$, the time scale is even smaller in our work, and self-capture remains negligible.


\end{enumerate}

\begin{table}[h!]
  \bigskip
    \centering\small\setlength\tabcolsep{4pt}
        \hspace*{-1cm}\begin{tabular}{l  | c c c | c c c c}
           \toprule
Benchmark & $m_X/\text{TeV}$ & $\alpha_X$ & $m_\phi/\text{MeV}$& $\epsilon^{-2}r_\text{th}/\lambda_\text{ion}$ & $\epsilon^{-2} r_\text{th}/\lambda_{\text{L}\alpha}$ & $\epsilon^{-2}r_\text{th}/\lambda_\text{C}^{(X)}$ &  $\epsilon^{-2}R_\odot/\lambda_\text{C}^{(e)}$ \\ [0.5ex] 
 \midrule
 \textbf{B1} & $0.4$ & $0.011$ & $10$  & $2 \cdot 10^{-22}$ & $5\cdot 10^{-24}$ & $2\cdot 10^{-28}$ & $6 \cdot 10^8$ \\ 
 \textbf{B2a} & $2$ & $0.040$ & $5$& $ 10^{-26}$ & $ 10^{-31}$ & $2\cdot 10^{-32}$ & $ 10^6$ \\
 \textbf{B2b} & $2$ & $0.040$ & $15$&  $2 \cdot 10^{-27}$ & $2\cdot 10^{-32}$ & $4\cdot 10^{-33}$ & $  10^6$ \\
 \textbf{B2c} & $2$ & $0.040$ & $70$&  $4 \cdot 10^{-29}$ & $8\cdot 10^{-34}$ & $ 10^{-34}$ & $ 10^6$ \\
 \textbf{B3a} & $139$ & $0.54$ & $2$&  $ 4\cdot 10^{-40}$ & $2\cdot 10^{-50}$ & $6\cdot 10^{-43}$ & $0.2$ \\
 \textbf{B3b} & $139$ & $0.54$ & $20$& $2 \cdot 10^{-41}$ & $2\cdot 10^{-51}$ & $4\cdot 10^{-44}$ & $0.2$\\
 \textbf{B3c} & $139$ & $0.54$ & $100$& $5 \cdot 10^{-43}$ & $4\cdot 10^{-53}$ & $ 10^{-45}$ & $0.2$ \\
           \bottomrule 
        \end{tabular}\hspace*{-1cm}
        \caption{Tabulated probabilities for a dark photon to ionize or excite ground state darkonium, as well as the likelihood of scattering on either DM or electrons before escaping from the Sun. The largest numbers appear in the last column, i.e. the largest effect we have neglected is scattering on electrons. Once the numbers are scaled by the relevant $\epsilon^2$ (typically around $10^{-20}$), scattering on electrons inside the Sun also become highly unlikely.}
\label{Tab: Probabilities} 
\end{table}

\section{Benchmark values}
\label{Sec: Benchmark summary appendix}
\begin{sidewaystable}[h!]
  \bigskip
    \centering\small\setlength\tabcolsep{4pt}
        \hspace*{-1cm}\begin{tabular}{l  | c c c | c c c c  c  c  c c c c c c }
           \toprule
             Benchmark & $m_X/\text{TeV}$ & $\alpha_X$ & $m_\phi/\text{MeV}$ & $\Delta$/GeV & $n_\text{max}$ & $\tilde{n}$ &  $C_\text{cap}/(\epsilon^2/\text{s})$ & $\tau_X/(\text{yr}/\epsilon)$ & $\sigma_\text{rec}/\sigma_\text{ann}$ & $N_X^\text{ss}/\epsilon$ & $N_o^\text{ss}/\epsilon^2$ & $N_p^\text{ss}/\epsilon^2$ & $r_\text{th}/R_\odot$ & $\epsilon_\text{min} \cdot 10^{10}$ & $\epsilon_\text{max} \cdot 10^{10}$ \\
           \midrule
              \textbf{B1} & $0.4$ & $0.011$ & $10$ & $0.012$ & $1$ & $1$ & $1.2\cdot 10^{39} $ & $0.0037$ & $3.1$ & $1.4\cdot 10^{44}$ & $1.0\cdot 10^{25}$ & $4.5 \cdot 10^{21}$ & $0.0054$ & $1.6 $ & $3.3 $  \\
              \textbf{B2a}  & $2$ & $0.040$ & $5$ & $0.80$ & $12$ & $7$ & $3.8\cdot 10^{38} $ & $0.00048$ & $13.5$ & $5.9\cdot 10^{42}$ & $3.6\cdot 10^{20}$ & $5.7 \cdot 10^{17}$ & $0.0024$ & $2.3 $ & $3.1 $ \\
              \textbf{B2b} & $2$ & $0.040$ & $15$& $0.80$ & $7$ & $5$ & $8.7\cdot 10^{37} $ & $0.0011$ & $10.8$ & $3.1\cdot 10^{42}$ & $8.0\cdot 10^{19}$ & $1.3 \cdot 10^{17}$ & $0.0024$ & $1.3$ & $3.5 $ \\
              \textbf{B2c} & $2$ & $0.040$ & $70$& $0.80$ & $3$ & $3$ & $3.0\cdot 10^{36} $ & $0.0071$ & $8.8$ & $6.8\cdot 10^{41}$ & $2.7\cdot 10^{18}$ & $4.4 \cdot 10^{15}$ & $0.0024$ & $0.62$ & $11 $ \\
              \textbf{B3a} & $139$ & $0.54$ & $2$& $10133$ & $2251$ & $216$ & $2.1\cdot 10^{36} $ & $0.000091$ & $30.3$ & $5.9\cdot 10^{39}$ & $4.8\cdot 10^{9}$ & $1.0 \cdot 10^{8}$ & $0.00029$ & $3.7$ & $8.1$ \\
              \textbf{B3b} & $139$ & $0.54$ & $20$& $10133$ & $712$ & $100$ & $1.4\cdot 10^{35} $ & $0.00038$ & $25.7$ & $1.7\cdot 10^{39}$ & $3.2\cdot 10^{8}$ & $7.0 \cdot 10^{6}$ & $0.00029$ & $1.2$ & $10 $ \\
              \textbf{B3c} & $139$ & $0.54$ & $100$& $10133$ & $318$ & $59$ & $3.6\cdot 10^{33} $ & $0.0025$ & $22.6$ & $2.8\cdot 10^{38}$ & $8.1\cdot 10^{6}$ & $1.8 \cdot 10^{5}$ & $0.00029$ & $0.52$ & $49 $ \\

           \bottomrule 
        \end{tabular}\hspace*{-1cm}
        \caption{Summary table of key benchmark values. The values for $n_\text{max}$ and $\tilde{n}$ have been rounded down to nearest integer.} 
        \label{Tab: Benchmark summary table}
\end{sidewaystable}

\clearpage

\section{Self-interaction and Bound state formation} \label{App: Schrodinger Eq}
In this section, we present how we numerically obtain the cross sections for self-interaction ($XX$, $X\bar{X}$, $\bar{X}\bar{X}$) and bound state formation ($X\bar{X}\to D$). We will use a non-relativistic approach where the scattering is described by a Yukawa potential\\
\beq
V(r) = \pm \frac{\alpha_X}{r} e^{-m_\phi r} \, ,
\eeq
where the plus (minus) is for repulsive (attractive) interactions. 
When comparing with constraints on self-interaction from DM distributions, the relevant quantity is the transfer cross section\\
\beq
\sigma_T =  \int d\Omega \, (1-\cos\theta) \,\frac{d\sigma}{d\Omega}.
\eeq
Analytical results are known within the Born approximation and the classical regime. In between these regions, both quantum mechanical and non-perturbative effects are important. We therefore have to solve the Schr\"odinger equation and use partial wave analysis to  express the transfer cross section as a sum over partial waves $\ell$\\
\beq\label{Eq: SI cross}
\frac{\sigma_T k^2}{4\pi} =  \sum_{\ell = 0}^{\infty} (\ell + 1) \sin^2 (\delta_{\ell+1} - \delta_\ell),
\eeq
where $\delta_\ell$ is the phase shift for the partial wave $\ell$ and $k=\mu v $ with $v$ the relative velocity and $\mu=m_X / 2$ the reduced mass. We obtain the phase shifts $\delta_\ell$ by solving the Schr\"odinger equation for the radial wave function $R_\ell (r)$ for the reduced two-particle system,\\
\beq\label{Eq: radial SE}
\frac{1}{r^2} \frac{d}{dr} \Big( r^2 \frac{d R_{\ell}}{dr} \Big) + \Big( k^2 - \frac{\ell (\ell + 1)}{r^2} - 2\mu V(r) \Big) R_\ell = 0
\eeq
and match $R_\ell (r)$ to its asymptotic solution\\
\beq
\lim_{r \to \infty} R_\ell(r) \propto  j_\ell(kr) \,  \cos\delta_\ell -  n_\ell(kr) \, \sin\delta_\ell \, ,
\eeq
expressed in terms of spherical Bessel (Neumann) functions $j_\ell$ ($n_\ell$). \\
\\
Within the dipole approximation, the cross section for bound state formation is given by \cite{An:2016gad} \\
%
\beq\label{Eq: BSF cross}
(\sigma v)_{\rm B} = \frac{\alpha_D}{3 \pi} \sum_{n, \ell} \left( \omega_{n\ell}^2 + \frac{1}{2} m_\phi^2 \right) \sqrt{\omega_{n\ell}^2 - m_\phi^2} \left[ \ell \left|\int dr r^3 \tilde{R}_{n\ell}R_{\ell-1} \right|^2 +(\ell+1) \left|\int dr r^3 \tilde{R}_{n\ell}R_{\ell+1} \right|^2 \right] \ ,
\eeq
where $R_\ell (r)$ is the radial wavefunction of the incoming state, satisfying Eq.~\eqref{Eq: radial SE}, while $\tilde{R}_{n\ell}$ is the radial wavefunction of the $(n\ell)$'th bound state of the Yukawa potential. The quantity $\omega_{n\ell} = E_{n\ell} + k^2/(2\mu)$ is the sum of the binding energy of the $(n\ell)$'th bound state and the kinetic energy of the incoming state.  In general, the binding energy of the Yukawa potential depends on both $n$ and $\ell$. However, since the size of those bound states, which are deep enough to emit an on-shell dark photon in their formation, is much smaller than $1/m_\phi$, we will use the following approximations:
\beq
E_{n\ell} \simeq E_{n} = \frac{\alpha_{X}^2 \mu}{2 n^2} \quad  \quad \tilde{R}_{n\ell} \simeq \tilde{R}_{n\ell}^{\text{Coulomb}}\, ,
\eeq
with 
\beq
\tilde{R}_{n\ell}^{\text{Coulomb}}(r) = \frac{2}{n^{\ell+2} (2\ell+1)!} \sqrt{\frac{(n+\ell)!}{(n-\ell-1)!}}\ \frac{(2r)^\ell}{a_0^{\ell+3/2}} e^{-\left( r/{n a_0}\right)} F_1 \left( 1+\ell -n, 2+2\ell, \frac{2 r}{n a_0} \right) \ ,
\eeq
where $a_0=1/(\alpha_X \mu)$ is the Bohr radius and $F_1$ is the Krummer confluent hypergeometric function. 

In order to solve the problem numerically, we introduce dimensionless quantities.
\beq
\chi_\ell \equiv r R_\ell \, , \quad x \equiv \alpha_X m_X r  \, , \quad a \equiv \frac{v}{2\alpha_X} \, , \quad b \equiv \frac{\alpha_X m_X}{m_\phi} \; ,
\eeq
and rewrite the Schrodinger equation as
\beq\label{Eq: dimless radial}
\left( \frac{d^2 }{d x^2} +  a^2 - \frac{\ell(\ell+1)}{x^2} \pm \frac{1}{x} \, e^{-x/b} \right) \chi_\ell(x) = 0 \; .
\eeq
\begin{figure}[h!]
	\centering
	\includegraphics[width=.6\textwidth]{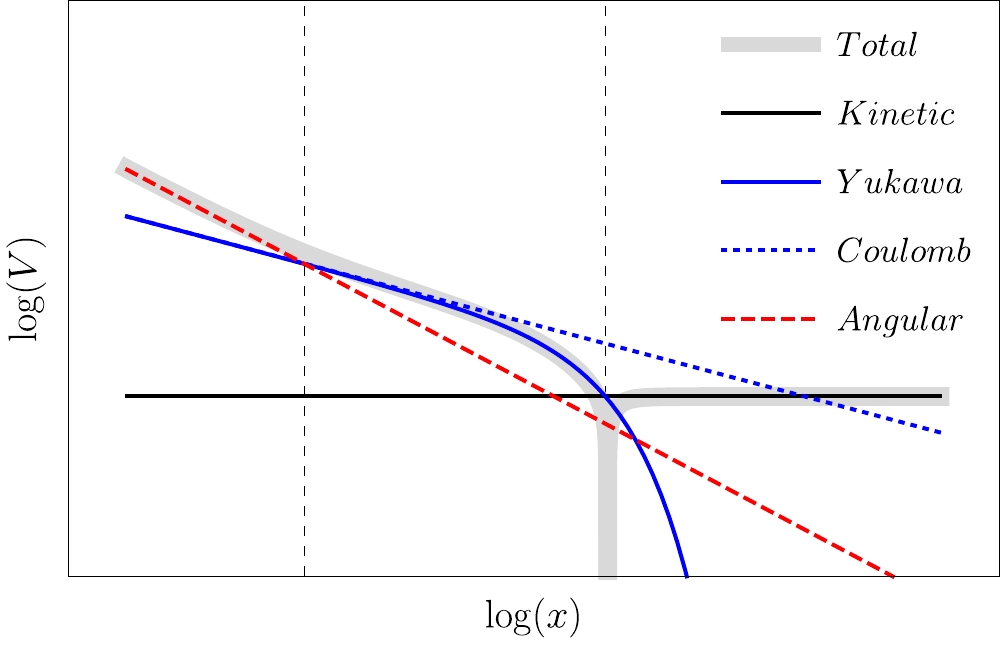}
	\caption{Sketch of the contributions to the curvature of the solution.}
	\label{fig7}
\end{figure}
We then use a slightly modified version of the numerical procedure in~\cite{Tulin:2013teo} to solve for $ \chi_\ell(x)$ (Step 1 and 2) and extract $\delta_\ell$ (Step 3) to be able to do the sum in Eq.~\ref{Eq: SI cross} (Step 4) and the integration in Eq.~\ref{Eq: BSF cross} (Step 6):
\begin{enumerate}
	\item \textit{Initial Conditions}  For \mbox{$x_i \ll b, (\ell+1)/a$}, Eq.~\eqref{Eq: dimless radial} is dominated by the angular momentum term, see Fig.~\ref{fig7}, and the solution $\chi_\ell(x) \propto x^{\ell +1}$. We thus impose as initial condition: 
	\beq
	\chi_\ell(x_i) = 1 \quad \quad \chi_\ell^\prime(x_i) = (\ell+1)/x_i\, .
	\eeq
	For $x_i \ll 1$ this is also true for $\ell=0$. The overall normalization is  fixed in step 5. 
	\item \textit{Radial Solution} We solve Eq.~\ref{Eq: dimless radial} numerically from $x_i$ to $x_e$. The end point $x_e$ is determined so we can afterwards  perform the steps \textit{matching}, \textit{normalization} and \textit{integration}.   The matching point $x_m$ is determined by the condition $a^2 \gg \exp(-x_m/b)/x_m$, where the potential term is suppressed compared to the kinetic term (see Fig.~\ref{fig7}). For normalization we need a few free oscillations, $x_{norm} > 2 \pi/a$, and we need to be able to do the  integrals of Eq.~\ref{Eq: BSF cross}, $x_{int} > 1/m_{\phi}$. 
	\item \textit{Matching} At $x=x_e \ge x_m$, we match $\chi_\ell$ (and its first derivative) onto the asymptotic solution, given by
	\beq
	\chi_\ell(x) \propto x \, e^{i \delta_\ell} \big(\cos\delta_\ell \, j_\ell(a x) - \sin \delta_\ell \, n_{\ell}(a x) \big) \; . \label{asympsol}
	\eeq
	Inverting Eq.~\eqref{asympsol}, the phase shift is given by
	\beq
	\tan \delta_\ell = \frac{a x_e \, j^\prime_\ell(a x_e) - \beta_\ell \, j_\ell(a x_e) }{a x_e \, n^\prime_\ell(a x_e) - \beta_\ell \, n_\ell(a x_e) } \; , \quad \beta_\ell = \frac{x_e \chi_\ell^\prime(x_e)}{\chi_\ell(x_e)} - 1,
	\eeq
	in terms of our numerical solution for $\chi_\ell$ at $x_e$.  The numerical method makes an initial guess for $(x_i,x_m)$ and computes $\delta_\ell$, and then successively decreases (increases) $x_i$ ($x_m$) until $\delta_\ell$ converges at $1\%$.
	\item \textit{Summation} We compute $\sigma_T$ by summing Eq.~\ref{Eq: SI cross} over $\ell$, truncating at $\ell_{\rm max}$.  We iterate $\ell_{\rm max}$ until $\sigma_T$ converges to $1\%$ and $\delta_{\ell_{\rm max}} < 0.01$ through ten successive iterations.
	\item \textit{Normalization} We normalize the solution $\chi_\ell(x)$ such that it has the same amplitude as the $\ell$-th partial wave for $x_e>x>x_m$.
	\item \textit{Integration} We perform the integrals in Eq.~\ref{Eq: BSF cross} for all the energy levels satisfying $E_n>m_\phi$.
\end{enumerate}

\end{document}